\documentclass[pra,twocolumn,graphicx,amssymb,floatfix]{revtex4}

\usepackage{graphicx}
\usepackage{amsmath}

\begin{document}

\title{Quantum Theory and Determinism}

\author{L. Vaidman}
\affiliation{ Raymond and Beverly Sackler School of Physics and Astronomy\\
 Tel-Aviv University, Tel-Aviv 69978, Israel}

\begin{abstract}
Historically, appearance of the quantum theory led to a prevailing view that Nature is indeterministic. The arguments for the indeterminism and proposals for indeterministic and deterministic approaches are reviewed. These include collapse theories, Bohmian Mechanics and the many-worlds interpretation. It is argued that ontic interpretations of the quantum wave function provide simpler and clearer physical explanation and that the many-worlds interpretation is the most attractive since it provides a deterministic and local theory for our physical Universe explaining the illusion of randomness and nonlocality in the world we experience.
\end{abstract}

\maketitle
\section{Introduction}

Quantum theory and determinism usually do not go together. A natural combination is quantum theory and randomness. Indeed, when in the end of 19th century physics seemed to be close to provide a very good deterministic explanation of all observed phenomena, Lord Kelvin identified ``two clouds'' on ``the beauty and clearness of the dynamical theory''. One of this ``clouds'' was the quantum theory which brought a consensus that there is randomness in physics. Recently we even ``certify'' randomness using quantum experiments \cite{Piron}.

I do not think that there is anything wrong with these experiments. They create numbers which we can safely consider ``random'' for various cryptographic tasks. But I feel  that we should not give up the idea that the Universe is governed by a deterministic law. Quantum theory is correct, but determinism is correct too. I will argue that the quantum theory of the  wave function of the Universe is a very successful deterministic theory fully consistent with our experimental evidence. However, it requires accepting that the world we experience is only  part of the reality and there are numerous parallel worlds. The existence of parallel worlds allows us to have a clear deterministic and local physical theory.

Before presenting this view I review how quantum theory led to believe that Nature is random. I give a critical review of attempts to construct theories with randomness underlying quantum theory. I discuss modifications of the standard formalism suggesting physical mechanisms for collapse. Then I turn to options for deterministic theories by discussing Bohmian mechanics and its variations, in particular a many Bohmian worlds proposal. Finally, I present the many-worlds interpretation and explain how one can deal with its most serious difficulty, the issue of probability.

\section{Determinism}

In my entry on the Many-Worlds Interpretation (MWI) \cite{MWISEP} I wrote that we should prefer the MWI relative to some other interpretations because it removes randomness from quantum mechanics and thus allows physics to be a deterministic theory. Last year I made a revision of the entry which was refereed. One of the comments of the referee was: ``Why I consider the fact that MWI is a deterministic theory a reason for believing it?'' I thought it is obvious: a theory which cannot predict what will happen next given all information that exist now, clearly is not as good as a theory which can.

It seems that in the end of the 19th century a referee would not ask such a question.
The dominant view then was that physics, consisted of Newton's mechanics and Maxwell's electrodynamics, is a deterministic theory which is very close to provide a complete explanation of Nature. Most scientists accepted a gedanken possibility of existence of ``Laplacean Demon'' \cite{Lapl1}:
\begin{quote}
 We may regard the present state of the Universe as the effect of its past and the cause of its future. An intellect which at a certain moment would know all forces that set nature in motion, and all positions of all items of which nature is composed, if this intellect were also vast enough to submit these data to analysis, it would embrace in a single formula the movements of the greatest bodies of the Universe and those of the tiniest atom; for such an intellect nothing would be uncertain and the future just like the past would be present before its eyes.

 \hfill  (Laplace, 1814)
\end{quote}
The idea of determinism has ancient roots \cite{Leuc}:
\begin{quote}
 Nothing occurs at random, but everything for a reason and by necessity.

 \hfill  (Leucippus, 440 BCE)
\end{quote}
Obvious tensions with the idea of a free will of a man or of a God led many philosophers to analyze this question. Probably the most clear and radical position was expressed by Spinoza \cite{Spin}:
\begin{quote}
 In nature there is nothing contingent, but all things have been determined from the necessity of the divine nature to exist and produce an effect in a certain way.

 \hfill  (Spinoza, 1677)
\end{quote}
Spinoza slightly preceded Leibniz who forcefully defended the Principle of Sufficient Reason \cite{Leib}:
\begin{quote}
 Everything proceeds mathematically ... if someone could have a sufficient insight into the inner parts of things, and in addition had remembrance and intelligence enough to consider all the circumstances and take them into account, he would be a prophet and see the future in the present as in a mirror.

 \hfill  (Leibniz, 1680)
\end{quote}
Hundred years ago Russell mentioned similar views, but already had some doubts \cite{Russ}:
\begin{quote}
 The law of causation, according to which later events can theoretically be predicted by means of earlier events, has often been held to be {\it a priori}, a necessity of thought, a category without which science would not be possible. These claims seem to me excessive. In certain directions the law has been verified empirically, and in other directions there is no positive evidence against it. But science can use it where it has been found to be true, without being forced into any assumption as to its truth in other fields. We cannot, therefore, feel any {\it a priori} certainty that causation must apply to human volitions.

 \hfill (Russell, 1914)
 \end{quote}
It was quantum theory which completely changed the general attitude. But the founders of quantum mechanics did not give up the idea of determinism easily. Schr\"odinger, Plank, and notably Einstein with his famous dictum: ``God does not play dice'', were standing against indeterminism. Earman, a contemporary philosopher who spent probably more effort on the issue of determinism than anyone else, writes \cite{Earm}:
\begin{quote}
 ... while there is no a priori guarantee that the laws of the ideal theory of physics will be deterministic, the history of physics shows that determinism is taken to be what might be termed a `defeasible methodological imperative': start by assuming that determinism is true; if the candidate laws discovered so far are not deterministic, then presume that there are other laws to be discovered, or that the ones so far discovered are only approximations to the correct laws; only after long and repeated failure may we entertain the hypothesis that the failure to find deterministic laws does not represent a lack of imagination or diligence on our part but reflects the fact that Nature is non-deterministic. An expression of this sentiment can be found in the work of Max Planck, one of the founders of quantum physics: determinism (a.k.a. the law of causality), he wrote, is a ``heuristic principle, a signpost and in my opinion the most valuable signpost we possess, to guide us through the motley disorder of events and to indicate the direction in which scientific inquiry should proceed in order to attain fruitful results [Plank, 1932]''

 \hfill  (Earman, 1986)
\end{quote}
I do not see a ``failure to find deterministic laws of physics''. All physical laws I studied, except for the collapse of the wave function which has many other properties which suggest to rejected it, are deterministic. I think that the prevailing view of indeterminism in the last century is an accidental mistake of the evolution of Science, similar to ether hypothesis rejected hundred years ago.

\section{ Probabilistic theories}\label{secProb}

Laplace, the symbol of determinism in physics is also the founder of probability calculus \cite{Lapl2}. He denied that there is an objective probability. The foundation of the probability theory is a realistic and deterministic theory with agents which are ignorant about some of the ontology.

The interpretation of probability is still a very controversial subject. The leading role in it plays de Finetti who also forcefully claims that there is no such thing as probability. It is only an effective concept of an ignorant agent \cite{deFin}:
\begin{quotation}
My thesis, paradoxically, and a little provocatively, but nonetheless genuinely, is simply this:

\centerline{PROBABILITY DOES NOT EXIST}

The abandonment of superstitious beliefs about the existence of the Phlogiston, the Cosmic Ether, Absolute Space and Time, . . . or Fairies and Witches was an essential step along the road to scientific thinking. Probability, too, if regarded as something endowed with some kind of objective existence, is no less a misleading misconception, an illusory attempt to exteriorize or materialize our true probabilistic beliefs.

 \hfill
(de Finetti, 1970)
\end{quotation}

 The program of presenting quantum theory as an objective probability theory cannot use classical probability theory since it assumes underlying definite values unknown to some agents. These definite values can be considered as ``hidden variables''. There are many limitations of the type of possible hidden variables, so it was acknowledged that probability theory underlying quantum theory cannot be a classical probability theory \cite{Accardi,Frohl}. Spekkens, who introduced a toy model which serves as an important test bed for many attempts in this direction, understands it well \cite{Spek07}:
 \begin{quote}
 It is important to bear in mind that one cannot derive quantum theory from the toy theory, nor from any simple modification thereof. The problem is that the toy theory is a theory of incomplete knowledge about local and noncontextual hidden variables, and it is well known that quantum theory cannot be understood in this way.

 \hfill (Spekkens, 2007)
 \end{quote}

 The program was put on the map long ago by Birkhoff and von Neumann \cite{BivonNE}. (See Pitowsky \cite{Pit} for development and defense of this position.)  A significant effort to find quantum theory emerging from probability calculus is Quantum-Bayesian interpretation of quantum theory (or QBism)\cite{CFS,Qbist2}. It should be mentioned that these developments came from the feeling that quantum theory cannot be understood in another way\cite{CFS}:
 \begin{quote}
 In the quantum world, maximal information is not complete and cannot be completed.

 \hfill (Caves, Fuchs and Schack, 2002)
 \end{quote}
 All authors of this program write that the quantum theory is intrinsically probabilistic. The program made a lot of technical progress borrowing results from flourishing field of quantum information. Axiomatic approaches of Popescu and Rohrlich \cite{PR} and Hardy \cite{Ha5Ax} brought interesting results. An operational approach to quantum probability which was put forward by Davies and Lewis \cite{DaLe}, was developed today into ``Generalized Probability Theories,'' (see recent review by Janotta and Hinrichsen \cite{JaHi}) which combines probability theory with quantum logic, the program envisioned by Birkhoff and Neumann \cite{BivonNE} long ago. A quite different  quantum logic appears in the ``Consistent (decoherent) histories'' approach \cite{Grif84,GellHart1990,Grif2013} which  includes
  inherently stochastic processes under all circumstances, not just when measurements
take place. Another post-classical probability theory, a ``convex-operational approach'' is conceptually more conservative. It differs from classical probability only due to rejection of the assumption that all measurements can be made simultaneously \cite{BaWi}. More demanding is the graphical framework for Bayesian inference \cite{CoeSp} which suggests replacing probability distributions with density operators and an attempt to formulate quantum theory as a causally neutral theory of Bayesian inference \cite{LeiSp}.

In my view, all these approaches are notoriously difficult. I agree with the predictions of Fr\"ohlich and Shubnel \cite{Frohl} regarding a gedanken poll among twenty-five grown up physicists asked to express their views on the foundations of quantum mechanics. It is indeed an ``intellectual scandal'' that there is nothing close to a consensus regarding the meaning of the most successful physical theory we have. But I am one of those colleagues who are convinced that ``somewhat advanced mathematical methods'' are superfluous in addressing the problems related to the foundations of quantum mechanics, and I turn off when I hear an expression such as ``$C^\star$-algebra'' or ``type-III factor''. I feel very comfortable with my approach. And indeed, as Fr\"ohlich and Shubnel predicted, ``almost all of them are convinced that theirs is the only sane point of view'', I do not see any other reasonable option. If I shall see that my option, the MWI, fails, I might turn to studying the operator algebra seriously.

I am skeptical about possibility to build quantum theory as a variation of a probability theory because the probabilistic aspects are not central in quantum theory. The unprecedented success of quantum theory is in calculation of spectrum of various elements, explaining stability of solids, superconductivity, superfluidity, etc. According to the probability theory approach to quantum theory, it is stated that ``the quantum state is a derived entity, it is a device for the bookkeeping of probabilities '' \cite{Pit}. But, for deriving all the results I mentioned above, I need the quantum state. It leads to an explanation for almost everything we observe in a very elegant and precise way. I cannot imagine how to calculate, say, the spectrum of Hydrogen with probability distributions instead of the quantum state. I am not aware of any explanation of non-probabilistic results of the quantum theory from some probability theory.

\section{ Uncertainty principle}

Arguably, the most influential result for today's consensus, that quantum theory is not a deterministic theory, is the Heisenberg Uncertainty Principle. In 1927 Heisenberg \cite{Hei} proved that an attempt to measure position of a particle introduces uncertainty in its momentum and vice versa.
 \begin{equation}\label{dxdp}
\Delta x ~ \Delta p \geq \frac {\hbar}{2}.
 \end{equation}

 Today, a more common term is the Uncertainty Relations, because attempts to derive quantum formalism from the Uncertainty Principle failed until now. Heisenberg was vague regarding quantitative description of the uncertainty. This is apparently the reason for the recent controversy regarding experimental demonstration of Heisenberg Uncertainty relations \cite{Oza2003,Erhard,Rozema,Bush,Dressel-Nori}.
 Uncertainty Relations became a hot topic of a recent research not just due to this controversy. Newly developed language of quantum information provided a possibility for alternative formulations and derivations based on entropic uncertainty relations. This direction was pioneered a while ago by Deutsch \cite{Deu83} and development continues until today \cite{Friedland}.

The argument of Heisenberg for indeterminism was that determinism as a starting port has a complete description of a system at the initial time. Inability to prepare the system with precise position and momentum does not allow precise prediction of the future.
(In classical physics complete description of the system was a point in the phase space: position and momentum.)
Robertson Uncertainty Relations \cite{Rob} is a more general representation of uncertainty principle because it puts constraint on the product of uncertainties of any pair of variables $A,B$:
\begin{equation}\label{dadb}
\Delta A ~ \Delta B \geq \frac {1}{2}\left|\langle \psi|[A,B]|\psi\rangle \right|.
 \end{equation}
A more important property of the Robertson Uncertainty Relation is that it states not just that we have no means to prepare a system with definite values of variables corresponding to noncommuting operators, the quantum formalism does not have a description for this. The Heisenberg uncertainty, i.e. the absence of a preparation procedure for a system with well-defined values of noncommuting variables is necessary to avoid the contradiction with the Robertson Uncertainty relation.

The fact that the Robertson Uncertainty Relation depends on the quantum state of the system sometimes  considered as a weakness.
The entropic approach to the Uncertainty Relations starts from the assumption that we deal here with probability distributions of values of the observables and can have a form independent of a particular quantum state. However, for the analysis of questions of determinism, the uncertainty relations corresponding to all states do not add much. The question is: ``Do the variables have certain values in a particular situation?'' Thus, what is relevant is the original Robertson Uncertainty Relation, (see its arguably simplest derivation \cite{Vdadb}). It teaches us that the quantum formalism does not allow definite values of variables corresponding to noncommuting operators. And the formalism has many pairs of such variables starting from position and momentum.
In Schr\"odinger representation of quantum theory the wave function is the basic concept. Obviously, typical wave function in position representation does not provide a definite position of a particle.

Uncertainty relations tell us that observables cannot have definite values in the framework of the quantum theory. They do not rule out hidden variables theories underlying quantum mechanics completing it in such a way that the observables {\it do} have definite values. The minimal disturbance meaning of the uncertainty relation does not change it. Hidden variables theory which specifies a deterministic evolution of hidden variables is not ruled out.

\section{ Kochen-Specker theorem}

Another apparently strong argument for the indeterminism related to the noncommutativity of operators corresponding to quantum observables is the Kochen-Specker theorem \cite{KS}. The theorem shows that assigning definite values to a particular set of observables together with the assumption that these values can be measured such that the noncontextuality is respected, i.e. measurements of commuting observables do not change the measured values of each other, contradict predictions of the quantum theory. The original proof was very complicated and had a set of 117 projectors. A lot of effort has been made to simplify it \cite{Peres,KS18}, and recently a system of just 13 projectors has been found \cite{KS13}. These results prove that at least some of the variables in the set do not have definite values. It has been strengthened also by a possibility to specify exactly which is provably value indefinite \cite{Abbot}.

In spite of the difficulty with experimental demonstration of Kochen-Specker contextuality in experiments with finite precision \cite{Me,Kent}, Cabello \cite{Ca08} found a feasible proposal and experiments have been performed \cite{Cab1,Cab2,CSneutr,KSphott}.
The experiments confirm predictions of quantum theory.

Uncertainty relations taught us that in some cases observables cannot have definite values in the framework of quantum theory.
The Kochen-Specker Theorem goes beyond it and puts constrains on the hidden variables theories. There is no way to have a noncontextual theory with definite values for quantum observables. But it does not prove randomness. A hidden variable theory for deterministic outcomes of all possible measurement is not ruled out. In fact, contextuality of a hidden variable theory is not something counterintuitive. An example of a contextual hidden variable theory is Bohmian mechanics described in Section \ref{secBohm}. It demonstrates the contextuality of a spin measurement in a very simple and natural way. The ``context'' here is not a simultaneous measurement of another observable. The context is a detailed description of the measuring device which measures one observable.
The outcome of the measurement is deterministic, but the observable, e.g. the spin component, does not have a definite value.

\section{ The Einstein-Podolsky-Rosen argument} \label{secEPR}

The formalism of the quantum mechanics with its two basic concepts, namely  quantum states and quantum observables, is an indeterministic theory due to the uncertainty principle: in certain situations some observables do not have definite values, the outcome of their measurements is indeterminate before the measurement. This is not what was expected from a good scientific theory at that time, but it did not contradict anything. The Kochen-Specker theorem showed that if we want to complete the quantum theory to avoid indeterministic outcomes, the hidden variables have to be ``contextual'', but it did not show a particular reason to complete the quantum theory: there was no inconsistency between what the theory predicts and what can be measured. In 1935 Einstein, Podolsky, and Rosen (EPR) \cite{EPR} showed such a reason.

\begin{figure}
\begin{center}
 \includegraphics[width=4.9cm]{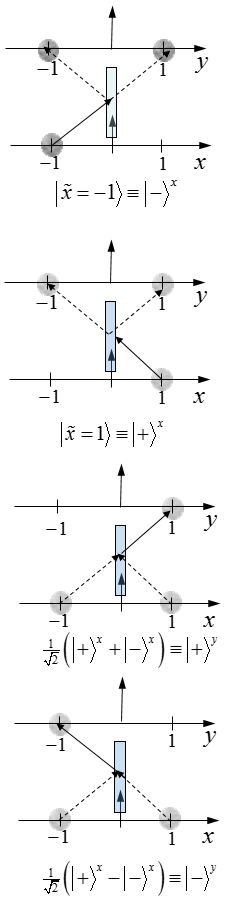}\end{center}
\caption{The first two figures show the eigen states of the variable $\tilde x$ which are the wave packets localized at points $\pm 1$ on the $x$ axis. The two figures in the bottom show the eigen states of the variable  $\tilde y$  which are superpositions of eigen states of $\tilde x$. They become localized wave packets on a parallel axis (which is named $y$) after passing through a vertical beam splitter.
The eigen states of  $\tilde x$ passing through the beam spitter become superpositions of the wave packets localized at $\pm 1$ on the $y$ axis.
}
\label{fig:1}
\end{figure}

The starting point of EPR was {\it locality}. They considered an entangled state which allowed measurement of variables of one system by measuring another system far away. They argued that these variables have to be definite before the measurement since, due to locality, the remote operation could not change it. If the values were not definite, they could not be present in the location of the remote measurement. The tension with quantum theory was that in the set of these variables there were noncommuting variables for which the uncertainty principle does not allow to assign definite values simultaneously. The EPR expected that completing the quantum theory with hidden variables which will assign definite values to these variables is possible. Naturally for EPR, the hidden variables were supposed to be local.
Following the analysis of Bohm and Aharonov \cite{BA}, Bell \cite{Bell1} showed that such local hidden variables do not exist.

The most vivid way to show this result is to consider a three-particle entangled state, named the Greenberger-Horne-Zeilinger state (GHZ) \cite{GHZ,Mer,VGHZ}. Let me present it in an abstract form considering the dichotomic variables $\tilde x$ and $\tilde y$ with the eigenvalues $\pm 1$ which are the counterparts of the spin variables of a spin$-\frac{1}{2}$ particle. The connections between eigenstates of $\tilde x$ and $\tilde y$ are:
\begin{equation}\label{xtilde}
  |\tilde y =\pm 1\rangle \equiv \frac{1} {\sqrt 2}|(|\tilde x=1\rangle \pm |\tilde x=-1\rangle).
\end{equation}

\begin{figure*}[h!tb]\begin{center}
 \includegraphics[width=0.7 \textwidth]{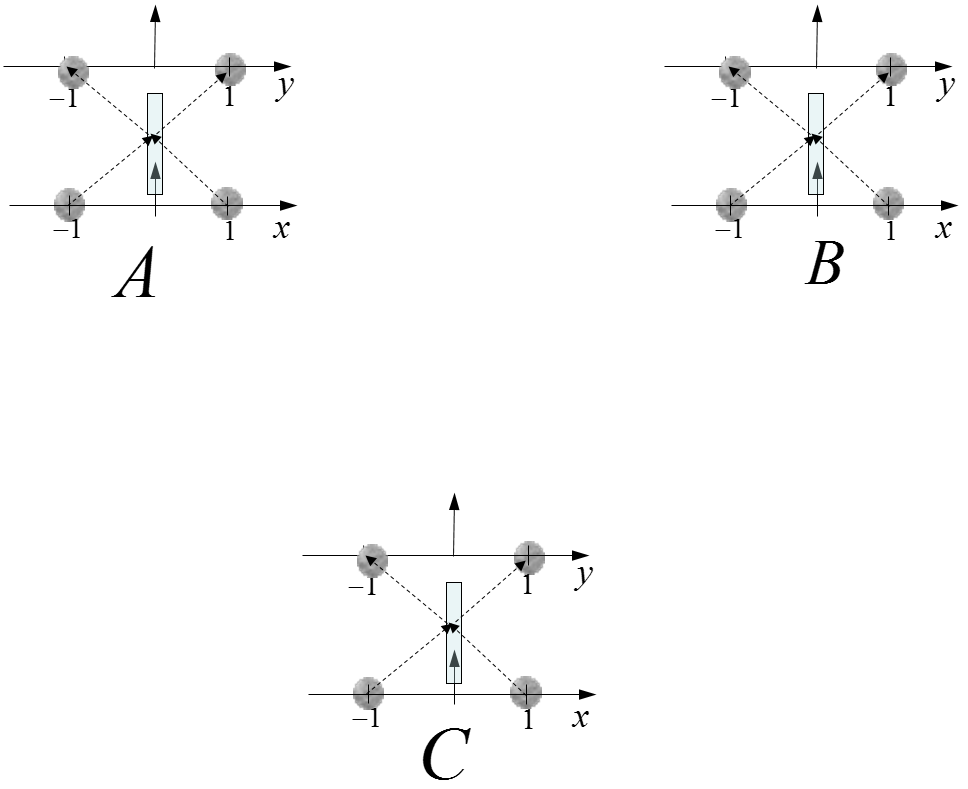}
\end{center}
\caption{The observers at sites $A$, $B$, and $C$ are asked simultaneously to measure $\tilde x$ or  $\tilde y$ of their particles. This is just either testing the presence of the particle at the locations $ x =\pm 1$, or passing it through the beam splitter, and then testing the presence of the particle at the locations $ y =\pm 1$. Quantum theory predicts the correlations between the outcomes of these measurements given by equations (\ref{XXX}) and (\ref{XYY}). Equation (\ref{square}) shows that these correlations are inconsistent with the assumption that the outcomes of the measurements are predetermined.  }
\label{fig:2}       
\end{figure*}

The variables $\tilde x$ and $\tilde y$ have a simple physical realization for a photon, allowing actual experiments in a laboratory. The eigenstates $|\tilde x =\pm 1\rangle$ by definition are localized wave packets at locations $ x =\pm 1$, see Fig.~1. Thus, measurement of $\tilde x$ is just a measurement of position. In order to measure the variable $\tilde y$, the photon, i.e. its two wave packets, should be sent through a properly placed beam splitter and then the photon position should be measured. The setup should be arranged such that the superposition of the wave packets with the same phase will interfere constructively at $y=1$ and destructively at $y=-1$, see Fig.~1.

Let us shorten the notation: $|\tilde x =+ 1\rangle_A =| + \rangle^x_A$ etc.
 Then the GHZ state is:
\begin{eqnarray}\label{GHZ1}
 \nonumber
 |GHZ\rangle\equiv \frac{1} {2} [ &|-\rangle^x_A&|+\rangle^x_B|+\rangle^x_C+|+\rangle^x_A|-\rangle^x_B|+\rangle^x_C\\
&+&|+\rangle^x_A|+\rangle^x_B|-\rangle^x_C-|-\rangle^x_A|-\rangle^x_B|-\rangle^x_C ].
\end{eqnarray}
All terms of the GHZ state fulfill the property:
\begin{equation}\label{XXX}
\tilde x_A~\tilde x_B~\tilde x_C=-1.
\end{equation}
If we change the basis in two sites, say $B$ and $C$, the GHZ state can be written in the form:
\begin{eqnarray}\label{GHZ2}
\nonumber
|GHZ\rangle = \frac{1} {2} [ |&-&\rangle^x_A|+\rangle^y_B|-\rangle^y_C+|-\rangle^x_A|-\rangle^y_B|+\rangle^y_C\\
&-& |+\rangle^x_A|-\rangle^y_B|-\rangle^y_C+|+\rangle^x_A|+\rangle^y_B|+\rangle^y_C ].
\end{eqnarray}
Similar expression can be obtained for changing the basis in other pairs of sites, so the GHZ state fulfills also the relations:
\begin{eqnarray}\label{XYY}
\nonumber
\tilde x_A~\tilde y_B~\tilde y_C&=&1, \\
\tilde y_A~\tilde x_B~\tilde y_C&=&1 , \\
\nonumber \tilde y_A~\tilde y_B~\tilde x_C&=&1.
\end{eqnarray}
When the system is in the GHZ state, we can measure the local variables $\tilde x$ and $\tilde y$ at each site by measuring the variables at other sites, see Fig.~2. The EPR argument tells us then that these variables should have definite values, but relations (\ref{XXX}),(\ref{XYY}) tell us that this is impossible: just take the product of all equations and we get a contradiction:
 \begin{equation}\label{square}
\tilde x_A^2\tilde x_B^2\tilde x_C^2\tilde y_A^2\tilde y_B^2\tilde x_C^2=-1.
\end{equation}

If we want to believe that every measurement ends up with a single outcome, namely the outcome which we  observe, then the EPR-Bell-GHZ result forces us to reject either locality or determinism. There is no local hidden variable theory which is consistent with predictions of quantum mechanics.
Six variables $\tilde x_A,\tilde x_B,\tilde x_C,\tilde y_A,\tilde y_B,\tilde y_C$ cannot have definite values prior to measurement. They are all locally measurable by detection photons in $x=\pm1$ positions or, after passing the photon wave packets through beam splitters, in $y=\pm1$ positions. Assuming locality, i.e. that the outcome of these measurements in each site depends solely on what is in this site only, ensures indeterminism. Presence of instantaneous nonlocal actions might save determinism: the definite values of variables in various sites might be changed due to measurements in other sites. Note that giving up determinism does not save locality: particular outcomes at two sites make, instantaneously, the values at the third site definite.

Let us summarize what we have shown until now regarding a possibility to change quantum formalism to make it a deterministic theory. The Uncertainty Relations told us that we have to add hidden variables. Kochen-Specker theorem told us that the hidden variables should be contextual. The EPR-Bell-GHZ result proved that the hidden variables have to be nonlocal.

\section{The meaning of the wave function }

In most of our discussion above, except for Section \ref{secProb}, the wave function $\Psi$ was tacitly assumed as part of the ontology. The discussion was about values of variables which could not be definite, but the role of $\Psi$ was not questioned.
The wave function considered to be ontic, i.e. to be  a description of reality, in contrast to  epistemic, corresponding to our knowledge of reality.

Even in the classic paper on statistical interpretation \cite{Ballentine} and in the stochastic model of Nelson \cite{Nelson}, the ontological role of $\Psi$ was not denied. Recently, however, with a new trend to consider quantum theory as part of an information theory, possibilities of theories in which $\Psi$ was epistemic were extensively investigated. Somewhat ironically, what brought the epistemic interpretation of $\Psi$ to the center of attention was a negative result by Pusey, Barrett, and Rudolph (PBR)  \cite{PBR}. Previously, the question was: ``Are there hidden variables in addition to $\Psi$, now a central question became ``Can hidden variables replace $\Psi$?'' Can it be that the ontology is these hidden variables, and $\Psi$ is just an emergent phenomenon?

In the standard quantum mechanics $\Psi$ is identified with a set of preparation procedures (many different procedures correspond to the same $\Psi$). Another manifestation of $\Psi$ is the set of probabilities for outcomes of all possible measurements. Pure quantum mechanics tells us that the systems obtained by identical or even by different preparation procedures but of the same state, are identical.

Usually, the existence of hidden variables denies that for every single system prepared in a state $\Psi$, the outcomes of measurements are uncertain, (or at least denies that they are given by the quantum probability formula). Instead, the outcomes are fixed by the hidden variables. The presence of the same ontological $\Psi$ for every instance of corresponding preparation procedure is not denied, as it explains the future evolution of the system in case the measurement is not performed.

The epistemic $\Psi$ is the idea that every preparation procedure corresponding to $\Psi$ ends up with a particular ontological state $\lambda_i$ (the hidden variable) and only distribution of parameters $\lambda_i$ in an ensemble of identical preparations is what corresponds to $\Psi$. Different $\Psi$s correspond to different distributions. But then we can imagine that a quantum system might have the same ontic state $\lambda_i$ when prepared in different  wave functions   $\Psi$ and  $\Psi'$. PBR proved that it is not possible, i.e. that the overlapping distributions of hidden variables for different quantum states contradict predictions of quantum mechanics. Thus, the wave function must be ontic as well.

The PBR proof involved analysis of two quantum systems and  measurements of entangled states of the composite system.
They
assumed that if we have two systems, then their hidden variables not only specify the outcomes of separate measurements on each system, but also outcomes of measurements of entangled states of the composite system. One can imagine that there are separate ontic hidden variables for each pair or each set of systems specifying the outcomes of measurements performed on composite systems. This contradicts the PBR assumptions and it is not a particularly attractive proposal, but it is a possibility, so the PBR proof is not unconditional. Note a parallel between the PBR and Bell's result \cite{Bell1}: Bell showed that there are no hidden variables for each system explaining outcomes of local measurements performed on two quantum systems in an entangled state, while PBR showed that there are no hidden variables for each system explaining the outcomes of a  measurement  of a variable with entangled eigen states performed on  the two systems prepared in a product state.

 Hardy \cite{Hardy} proves inconsistency of the overlapping distributions based on another assumption which he named ``ontic indifference'': the operations which do not change $\Psi$ also do not change the underlying hidden variables (more precisely, there exist an implementation for operations on hidden variables with such a property).
Patra et al. \cite{PaMas1} reached similar conclusions assuming a certain ``continuity assumption'' according to which small enough change of $\Psi$ does not lead to the change of $\lambda_i$. An experiment \cite{Patraexpereiment} ruled out some of the epistemic models which predict deviations from the standard quantum theory.

Colbeck and Renner \cite{CR2} claimed to resolve the issue of the meaning of $\Psi$ based only on ``the assumption that measurement settings can be chosen freely'' (FR). Their conclusion is that ``a system's wave function is in
one-to-one correspondence with its elements of reality''. This conclusion is what I wish to obtain. However, unfortunately, I was not convinced by their formal arguments. The Colbeck and Renner result heavily relies on their previous claim that ``No extension of quantum theory can have improved predictive power'' \cite{CR1} based on the same FR assumption.
When I read this work, it looks circular. They assume that the present quantum theory is correct and that the extension is accessible. Then, if this extension can help to predict some outcomes, it  violates statistical predictions of quantum theory.
However, reading what they say about Bohmian Mechanics, I realized that their  whole approach to what is a scientific theory is different from what I consider here. They wrote that the Bohmian theory contradicts their  FR assumption of the free choice of measurement settings. As I explain in the next paragraph, the  measurement settings are free by definition. See Ghirardi and Romano \cite{GRFound} for the analysis of the Colbeck-Renner FR assumption and Laudisa \cite{Lad} for an illuminating discussion of Colbeck-Renner and other ``no-go'' quantum theorems (cf Bell's work on ``no-go'' quantum theorems \cite{Bell-no-go}).

Although what I analyze in this paper is supposed to explain everything, the whole physical Universe which includes even more than just one world we are aware of, my attitude is that it is enough to have a satisfactory theory for a closed system, for a small box with a few particles, for a room with an observer, for the planet Earth. The theory is supposed to answer correctly about the result of every experiment we imagine to perform on this system. In particular, any preparation, any intermediate disturbances, any intermediate and final measurements should be considered without constraints. We, outside the closed system, have a complete ``free choice'' of our actions. I am ready to extrapolate that if the experiments confirm all the results for which we are capable to calculate the predictions according to our theory, it also explains well our Universe in which there are no agents with ``free will'' to make the choice of various options in quantum experiments. Thus, I am not worried about ``The Free Will Theorem'' \cite{CoKo}.

As Harrigan and Spekkens admit \cite{HaSp}, today there is no serious candidate for an epistemic model of $\Psi$. Moreover, recent research, notably by Montina \cite{Montana} and very recently by Leifer \cite{Leifer} does not lead us to expect that the complexity of such a model will be smaller than that of the standard quantum theory. The large complexity of classical models underlying quantum seems to be necessary for explaining the huge difference between the power of a qubit versus the power of a bit for various information tasks \cite{GaHa,VaMit}. The wave function provides a simple and elegant explanation of these protocols. The strongest motivation for the ontology of $\Psi$ is one of the oldest: it provides the simplest explanation for the particle interference.

The work on protective measurements \cite{PMeas} also supports the ontological view of $\Psi$. The fact that we can observe the wave function (until today only in a gedanken experiment) of a single particle suggests that it is the ontological property of the particle. However, it is not a decisive argument since observation of the wave function requires long time interaction and ``protection'' of the state. It is possible that the protection procedure acts on the elements of reality $\lambda_i$ enforcing its motion such that its average during the time of protective measurement will create the shape of the wave function.
The chances to have such a mechanism seem slim. We already have a theory, and not a very simple one: the quantum theory describing unitary evolution of the wave function which makes specific predictions for the results of many possible measurements. The alternative theory should provide identical, or at least very similar, predictions and in many very different situations. It should work for all Hamiltonians in which the quantum wave we consider is one of the eigenfunctions. It should work also for all kinds of frequent projection measurements on the quantum state in question.

At the early days of quantum theory, the motivation for a search of an epistemic interpretation of quantum theory was a sentiment for classical physics explanation which was considered to be very successful before the quantum theory appeared. Also, the success of statistical mechanics in explaining thermodynamics suggested that a similar relation might exist between classical and quantum theories. Later, when the quantum theory explained with unprecedented precision the majority of observed physical phenomena, the main motivation for an epistemic interpretation was its simple and elegant explanation of the collapse of $\Psi$ in quantum theory, ``an ugly scar on what would be a beautiful theory if it could be removed'' \cite{Got}. Nowadays, the explosion of the works analyzing epistemic approach \cite{Psi2,Psi3,Psi4,Psi5,Psi6,Psi7} which however, mostly produce negative results, can be explained by the development of quantum information theory which provides new tools which make these analyses possible. The ``positive'' result \cite{Lewis1}, showing a possibility of $\Psi$-epistemic model, seems to me not convincing because this proposal is too conspiratorial. I will turn now to the discussion  of interpretations in which $\Psi$ has ontological meaning, what is today frequently named as {\it ontic} interpretation of $\Psi$.

\section{Collapse models}

Quantum system, according to von Neumann \cite{vonNeuma}, evolves according to the Schr\"odinger equation between measurements at which it collapses to the eigenstate of the measured variable. The problem with this simple prescription is that there is no definition what is ``measurement'' \cite{againstmeasurBell}. The concept is frequently ``clarified'' by a statement that measurement happens when ``macroscopic'' measuring device makes a recording which only replaces one ill-defined concept by another. Von Neumann understood it well and he added a proof that for all practical purposes, i.e. for all observed experimental results, it is not important where exactly we place the ``cut'', the point when measurement really happens. While this proof allows us to use quantum theory for predicting results and building useful devices, it does not allow us to consider quantum theory as a description of Nature. I believe that the latter is not just philosophical and academic question. Answering it might lead to better ways for predicting results of experiments and designing useful devices.

At 1976 Pearle \cite{Pear1} proposed a mechanism for the physical collapse of $\Psi$ by adding a nonlinear term to Schr\"odinger equation. Ten years later Ghirardi, Rimini and Weber \cite{GRW} (GRW) proposed much simpler but {\it ad hoc} physical postulate which, followed support by Bell \cite{Bell2}, triggered an extensive development of the collapse program. The most promising direction is the Continuous Spontaneous Localization (CSL) models started by Pearle \cite{Pear2}, and significantly developed using the GRW ideas \cite{GirarPearle}. In the main approach, the wave function $\Psi$ is the ontology. Thus, the quantum interference is explained as in any other wave phenomena. The evolution of $\Psi$ is, however, not unitary. The Schr\"odinger evolution is modified by some explicitly random element. Randomness is the property of all collapse proposals. The ongoing research is to add some physical explanation why the random collapse occurs. Note gravitation induced collapse proposals by Diosi and Penrose \cite{Dio,Penr}.

In the GRW proposal, the wave function of every particle is multiplied at a random time, on average once in $\tau=10^{8}$ years, by a Gaussian with a width of $d=10^{-5}$cm. The location of the center of the Gaussian is chosen randomly but in proportion to $|\Psi|^2$. In experiments with a few particles these GRW ``hits'' usually are not observable, because they are very rare. If we have a well localized macroscopic body, these hits also will not do much. Electrons in atoms are localized with a width of the order of $10^{-8}$cm, so multiplication by a much wider Gaussian does not modify the wave function significantly. The situation is different if a macroscopic body, say a pointer, is in a superposition of being in two well separated places. Very fast, at least one particle of the body will be multiplied by the GRW Gaussian which will localize, due to the entanglement between the particles in the pointer, the whole pointer to one position.

The GRW proposal is phenomenological, ungrounded in more fundamental physics, as are the CSL models with recent arguments for  uniqueness \cite{BDH,PearleY80}. But even the GRW program of the dynamical collapse is a physical theory which is a candidate for a complete description of Nature. We do not need any additional clarification of the concepts: What is a measurement? What is macroscopic? The theory tells us when it is probable that the collapse takes place and this can be considered as the time when a quantum measurement procedure ends with a particular result.

In a setup of a quantum measurement which shows the outcome by positioning a pointer having a macroscopic number of atoms, the theory predicts a reasonable behavior. The ``cut'' between the quantum superposition stage and the collapsed state is in the reasonable place: the pointer stays in a superposition of macroscopically different states for an extremely short time.
Albert and Vaidman \cite{AlV} noted that in some other setups the ``cut'' placed by the GRW might be far beyond the time one might expect. Although ``macroscopic'' is not rigorously defined, it seems natural to consider $10^{10}$ atoms as a macroscopic object. So, we would expect from a collapse model to cause a very fast reduction of a superposition of states in which all atoms are excited and in which all atoms are in their ground state. However, if the difference in the size of the electron cloud in the excited and the ground states is much smaller than $10^{-5}$cm, then the GRW hit does not lead to a significant change of the wave function of an individual electron, and thus does not lead to the collapse of the superposition. Completely different pictures on a screen ``drawn'' by areas of such excited  atoms are not ``macroscopically different'' according to the GRW precise definition.

\begin{figure}
\begin{center}
 \includegraphics[width=7.2cm]{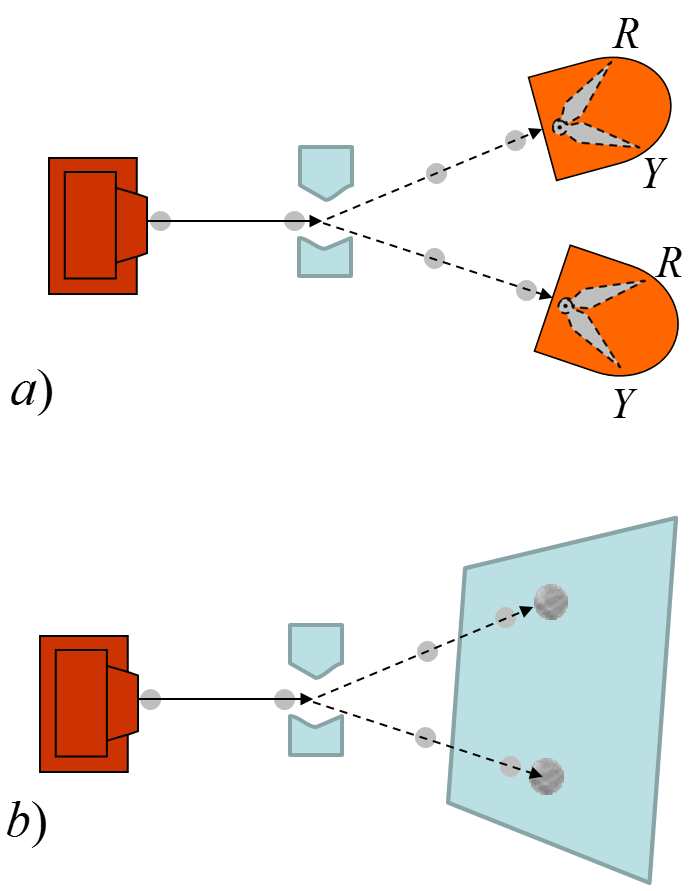}\end{center}
\caption{ a). The Stern-Gerlach experiment with  pointer displays. $R$ refers to READY position and $Y$ to YES position of the pointer which signifies detection of the atom.
The superposition of the wave packets of the atom creates an entangled superposition of the two pointers which collapses almost to a product state due to the first GRW hit. b). The Stern-Gerlach experiment with a screen display. The superposition of the wave packets of the atom creates an entangled superposition of the internal states of a macroscopic number of  atoms on the screen. This superposition does not collapse due to the GRW hits, so the definite result of the measurement will take place at a much later stage.}
\label{fig:3}
\end{figure}

Consider a Stern-Gerlach experiment measuring spin in the $z$ direction of a particle prepared with the spin ``up'' in the $x$ direction. First, consider a measuring device which shows the result using macroscopic pointer starting in $R$ (READY) position and ending up in $R$ or $Y$ (YES) position signifying detection of the particle, see Fig.~3a. The unitary evolution  leads to a superposition of two macroscopically different pointer positions, but then, the first GRW hit on one of the atoms of the pointer will localize the pointer to one position.
If, however, the detector is just a plate in which numerous atoms are excited around the place the atom hits the plate, the GRW hits will not eliminate the superposition of different sets of a macroscopic number of excited atoms, see Fig.~3b.

It has been shown \cite{GRWanswer} that this does not lead to an observable difference because the GRW mechanism collapses the wave function when the neurons of the visual cortex of the observer's brain transmit the signal. Still, it seems a weakness of the collapse model that in this example the von Neumann cut should be put so close to the human perception. We would expect that a physical theory will tell us that what we see is what is, but here we are told that we play an active role in creating the reality by our observation. In every quantum textbook we read that quantum measurements play an active role in forming the ``reality'', but it seems that the measuring devices should be responsible for this, not our brains.

There is another aspect in which the GRW collapse apparently provides less than the von Neumann collapse. According to the von Neumann, at the end of the measurement process, the part of the wave corresponding to the outcome which was not found, disappears completely. In the Stern-Gerlach experiment with a pointer display, Fig.~3a, after the measurement, there will be a quantum state describing the pointer just in one position. The process can be described as follows:
\begin{eqnarray}\label{vNpro}\nonumber
 \frac{1}{\sqrt{2}}(|{\uparrow}\rangle+|{\downarrow}\rangle)|R\rangle^{A}|R\rangle^{B}&\rightarrow& \\
 \rightarrow \nonumber \frac{1}{\sqrt{2}}(|{\uparrow}\rangle|Y\rangle^{A}|R\rangle^{B}&+&|{\downarrow}\rangle|R\rangle^{A}|Y\rangle^{B})\rightarrow \\ &\rightarrow& |{\uparrow}\rangle|Y\rangle^{A}|R\rangle^{B},
\end{eqnarray}
where $|Y\rangle^{A}$ signifies ``YES'', the state of the upper detector detecting the atom, $|R\rangle^{B}$ signifies the lower detector in the state ``READY'', etc.
The GRW provides similar, but not exactly the same evolution:
\begin{eqnarray}\label{GRWpro}
 \nonumber
 \frac{1}{\sqrt{2}}(|{\uparrow}\rangle+|{\downarrow}\rangle)|R\rangle^{A}|R\rangle^{B}&\rightarrow & \\ \nonumber \frac{1}{\sqrt{2}}(|{\uparrow}\rangle|Y\rangle^{A}|R\rangle^{B}&+&|{\downarrow}\rangle|R\rangle^{A}|Y\rangle^{B})\rightarrow \\
 \rightarrow \mathcal{N}(|{\uparrow}\rangle|Y\rangle^{A}|R\rangle^{B}&+&e^{-\frac{l^2}{2d^2}}|{\downarrow}\rangle|R\rangle^{A}|Y\rangle^{B}),
\end{eqnarray}
where $d=10^{-5}$cm is the GHZ parameter and $l\simeq 5$cm is the distance between the atom in the pointer which was ``hit'' by the GRW collapse mechanism in the READY and YES positions.

\begin{figure*}[h!tb]\begin{center}
 \includegraphics[width=0.75\textwidth]{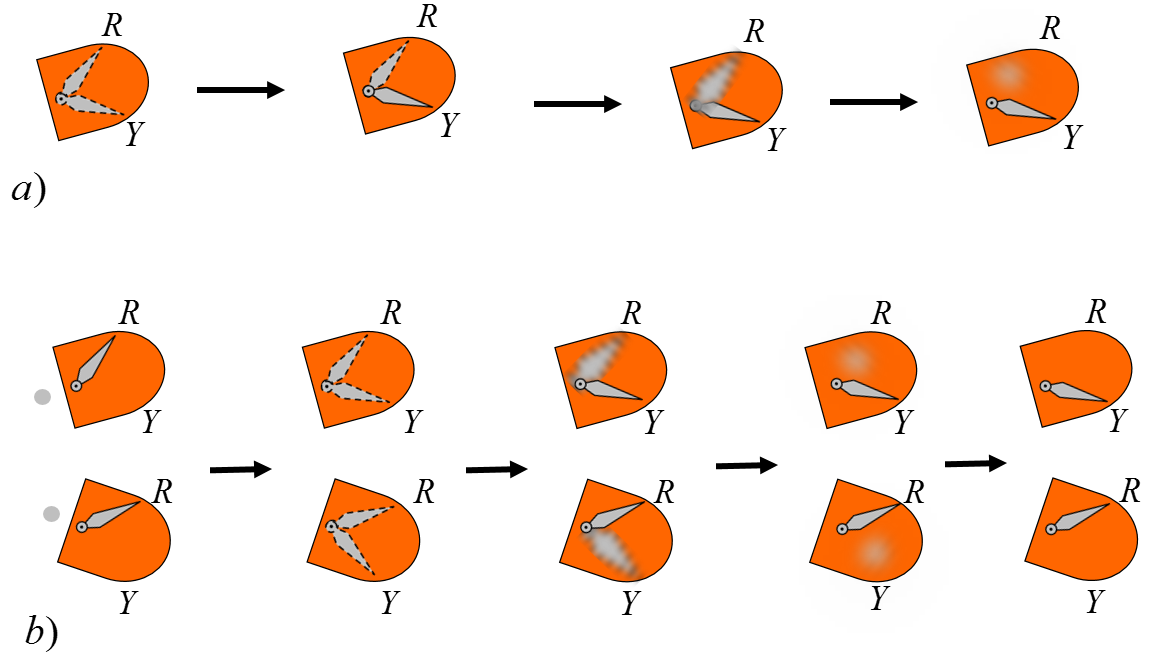}
\end{center}
\caption{a). The dynamics of the GRW collapse of the wave function of  the pointer in superposition. The first GRW hit collapses the amplitude of the tail pointer state to a very small number. The following hits quickly destroy the tail state of the pointer such that the pointer disintegrates. b). In the Stern-Gerlach experiment, Fig.~3a, the GRW hits lead to  a single stable structure of the pointer's states. ($R$ refers to READY position and $Y$ to YES position of the pointer which signifies detection of the particle.) }
\label{fig:4}       
\end{figure*}

This is the GRW tail problem \cite{tail1,tail2,tail3}. The wave function after the measurement according to the GRW is a superposition of the state corresponding to one outcome with an amplitude close to 1 and the state corresponding to the other outcome with an exponentially small amplitude. With time, this amplitude will be reduced very fast to even smaller values. In my view, as presented, it is still a weakness of the theory. The absolute value of the amplitude does not change the experience of the observers having such an amplitude. It seems that the GRW mechanism fails to perform the task it was constructed for: to eliminate all but one of the outcomes of a quantum measurement. True, it singles out one outcome as the only one having a large amplitude, but it apparently leaves the other outcomes coexisting. The GRW proponents have to explain: ``How do I know that I am not the one in the tail branch?''

A popular attempt to overcome this difficulty is to define a cutoff declaring that if the amplitude of a branch is smaller than some number, then this branch can be neglected. I do not find this proposal appealing.
I see, however, another resolution of the GRW tails problem. The GRW mechanism literary speaking ``kills'' the tail branches.

Consider the wave function of an electron of an atom in the pointer being in a superposition of two readings after the hit by a Gaussian, Fig.~4a. The center of the Gaussian will be at the location of an atom corresponding to one of the positions of the pointer. Since the width of the Gaussian is three orders of magnitude larger than the diameter of the electron wave function in the atom $D\simeq 10^{-8}$cm, the atom will not change its state.
On the other hand, the center of the Gaussian will be at a macroscopic distance $l$ from the nucleus of the atom in the other position of the pointer. The wave function of the electron in the atom $\psi(r)$ will be multiplied by $e^{-\frac{(r-l)^2}{2d^2}}$. It will not be just multiplied by a small number, it will be severely disturbed. Indeed, the part of the wave function which is far from the center will be reduced relative the part which is close to the center by the factor
\begin{equation}\label{factor}
e^{-\frac{(l+D)^2}{2d^2}}/e^{-\frac{l^2}{2d^2}}\simeq e^{-\frac{lD}{d^2}}\simeq 10^{-4}.
\end{equation}
Clearly, the atom will change its state and most probably will be ionized. Thus, the final term in (\ref{GRWpro}) describes well the situation after the first GRW hit up to a disturbed state of one atom in the tail of the wave function. But very soon, numerous other atoms in the tail wave function of the pointer will be severely disturbed, see Fig.~4a. After a short time the pointer will disintegrate in the tail wave function up to a situation that we should say that the pointer is not there anymore. Fig.~4b. shows the disappearance of the tails states of the detectors in the Stern-Gerlach experiment.

In the same way, the GRW mechanism in a situation of a human observer being in a superposition of macroscopically different positions immediately kills the person in all tail branches and reduces the wave function to a single stable branch of the person. There is no way to have conscious beings in more than one branch in the GRW approach. This, together with the smallness of tails (which is necessary for having our existence plausible) resolves the tail problem.

The collapse models are observably different from the standard quantum mechanics. The GRW collapse mechanism leads to a tiny energy non-conservation \cite{Pear3,Adl}. These effects were not found \cite{Col95,Bowm,Arndt}, so the original GRW proposal was ruled out, but some of its modifications and some CSL models are still possible \cite{Bassi}.

In recent years there have been proposed modifications in viewing the ontology of the collapse models. In addition to $\Psi$ the ``mass density'' \cite{GRWm} and ``flashes'' \cite{Tumu,EsfGis} that are responsible for the random collapses were suggested as ontological entities. Adding other ontological entities besides $\Psi$ seems to me (and to Albert \cite{AlbertWF}) unwanted. I view the strength of collapse models that they do not require anything like hidden variables to avoid plurality of worlds. We can consider GRW hitting process and the CSL  fluctuating field as part of the physical law and take $\Psi$ as the only ontology. (Surely it is less radical than considering the wave function of the Universe as a law of motion of Bohmian particles \cite{GoldZang}.)
 Apparently, the reason for introducing additional ontology is the difficulty to explain our experience based on $\Psi$ ontology only \cite{Maudlin}. In Sec \ref{secMWI} I will show how our experience {\it can} be explained in a satisfactory way based solely on $\Psi$, so it is not necessary to introduce the ``mass density''  or ``flashes'' ontology.

An interesting option to obtain an effective collapse without introducing any complicated dynamics is to accept that there are two ontological wave functions, the usual one evolving toward the future, specified by the boundary condition in the past, and another one evolving backward in time, specified by the boundary conditions in the future \cite{ABL,AGrus,CohAha}. The boundary conditions in the future should correspond to the results of all measurements in the Universe and thus replacing all the collapses which were supposed to happen. The outcomes of these measurements provide the real part of {\it weak values} \cite{AV90} of measured values which describe well the world we experience. It is apparently a consistent proposal, and the high price of a very complicated backward evolving wave function maybe reasonable for avoiding collapse. I think, however, that the multiple worlds are not as problematic as they are usually viewed, so the conspiracy of the specific backward evolving wave function is a too high price for avoiding parallel worlds \cite{tisy}.

\section{Bohmian mechanics}\label{secBohm}

By far the most successful hidden variables theory is the Bohmian mechanics. It is a deterministic theory capable to explain the appearance of probability. It reproduces all predictions of quantum theory but it also provides a convincing explanation of quantum peculiar phenomena. It provides a solution of quantum measurement problem reproducing an effective collapse of the wave function. It nicely demonstrates noncontextuality and beautifully explains the EPR-Bell-GHZ correlations. It is a candidate for a final theory of the world. One of the reasons for its popularity is that the Bohmian picture of reality is close to the Laplacian picture. The world is a collection of particles with well-defined trajectories. The law of evolution of Bohmian particles is more subtle than Newton's laws and it requires $\Psi$ which also obtains an ontological status.

\begin{figure*}[h!tb]\begin{center}
 \includegraphics[width=0.75\textwidth]{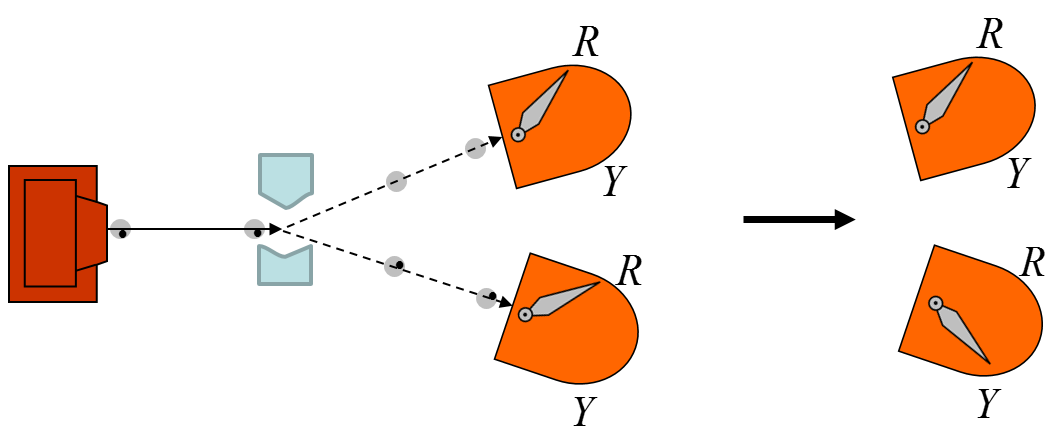}
\end{center}
\caption{When the standard quantum theory predicts an equal probability for the results of the Stern-Gerlach experiment, the Bohmian position in the lower half of the wave packet of the particle ensures that it will be detected by the lower detector. Depending on the type of the Stern-Gerlach magnet, this will correspond to the results: spin ``down'' or spin ``up''. ($R$ refers to READY position and $Y$ to YES position of the pointer which signifies detection of the particle.) }
\label{fig:5}       
\end{figure*}

 De Broglie \cite{Broglie28} was first to suggest a version of the Bohmian mechanics in 1927 but he changed his position, presenting a significantly different view later \cite{deBroglie1956}. Bohm \cite{Bohm52} made a clear exposition of the theory in 1952, although he never viewed it as a proposal for a final theory. For Bohm it was a way for developing a new and better approach. The motions of ``particles'' in the de Broglie and Bohm theories are identical, but the important difference between these formulations is that de Broglie used an equation for velocity (determined by the quantum state) as the guiding equation for the particle, while Bohm used the equation for acceleration (a la Newton) introducing a ``quantum potential'' (likewise determined by the quantum state). The version I find the most attractive was advocated by Bell \cite{bell2004speakable} and it is closer to de Broglie as the equation of motion is given in terms of the velocity.
An important aspect is that the only ``Bohmian'' variables in this approach are particle positions, and all other variables (e.g. spin) are described solely by the quantum state.

The ontology of the Bohmian mechanics consists of the wave function $\Psi$ and the trajectories of all particles in three-dimensional space. The quantum state evolves according to the Schr\"odinger equation (or, more precisely, according to its relativistic generalization) and it never collapses. It is completely deterministic, the value of the wave function at any single time determines the wave function at all times in the future and in the past.
 Every particle has a definite position at all times and its motion is governed in a simple deterministic way by the quantum state.
 The velocity of each particle at a particular time depends on the wave function and positions of all the particles at that time. Namely, the velocity of particle $i$ at time $t$ is given by:
 \begin{equation}\label{velo1}
\dot{\boldsymbol{\textbf{r}}}_{i}(t)=
\textbf{Im}\frac{\hbar}{m}\left.\frac{\mathbf{\boldsymbol{\Psi}}^{\dagger}\left(\boldsymbol{r}_{1},\ldots\boldsymbol{r}_{N},t\right)
\boldsymbol{\nabla}_{i}\boldsymbol{\Psi}\left(\boldsymbol{r}_{1},\ldots\boldsymbol{r}_{N},t\right)}{\boldsymbol{\Psi}^{\dagger}
\left(\boldsymbol{r}_{1},\ldots\boldsymbol{r}_{N},t\right)\boldsymbol{\Psi}\left(\boldsymbol{r}_{1},\ldots\boldsymbol{r}_{N},t\right)}\right|_{\boldsymbol{r}_{i}=
\boldsymbol{\textbf{r}}_{i}(t)}.
\end{equation}
(We use Roman text font for Bohmian positions and bold font to signify
that $\mathbf{\boldsymbol{\Psi}}$ is a spinor.)
 This velocity formula ensures that a Bohmian particle inside a moving wave packet which has group velocity $\bf{v}$ ``rides'' on the wave with the same velocity, $\dot{\mathbf{r}}=\mathbf{v}$.

It is postulated that the initial distribution of Bohmian positions of particles is according to  the Born rule, i.e. proportional to $|\Psi(\textbf{r})|^{2}$. Then, the guiding equation ensures that this Born law distribution will remain forever \cite{Golds}. In fact,  even if the Bohmian position of a particular system starts in a low probability region, due to interactions with other systems it will typically move to a high probability point very rapidly
\cite{PhysRev.89.458,Valentini2005}. Thus, postulating an initial Born rule distribution might not be necessary. However, some restrictions on the initial Bohmian position are unavoidable: it cannot be where $\Psi(\textbf{r})=0$. We have not seen deviations from the standard quantum theory, so I find it preferable to keep the initial Born distribution postulate in Bohmian theory. It is possible, however, that cosmological considerations might lead to dynamical origin of the Born rule \cite{Valentini2010}).

If we want to consider observables of quantum theory as basic concepts, the Bohmian mechanics demonstrates contextuality. When the wave function and the Bohmian positions of all particles are given, the outcomes of measurements of the observables might not be fixed. This statement does not contradict the determinism of the Bohmian mechanics: the outcome of every experiment is predetermined. However, different experimental setups for measurements of the same observable might lead to different observed values. The ``context'' here is not a simultaneous measurement of another observable as in Kochen-Specker theorem. The context is a detailed description of the measuring device which measures a single observable.

Consider a Stern-Gerlach measurement of a spin $z$ component of a spin$-\frac{1}{2}$ particle prepared initially ``up'' in the $x$ direction. Quantum mechanics predicts equal probability for ``up'' and ``down'' spin $z$ outcomes. In a Stern-Gerlach experiment the particle passes through an inhomogeneous magnetic field and the ``up'' and ``down'' components end up in different places, Fig 3. Assume that ``up'' spin goes to an upper place and ``down'' spin goes to the lower spot. Spin $x$ state is a superposition of spin $z$ ``up'' and ``down'' and thus, the quantum theory does not tell us where the particle will end.

The Bohmian quantum mechanics is a deterministic theory and given the quantum state and the initial Bohmian position, the future is definite. The analysis here is  simple \cite{gili}, Fig.~5. The Bohmian position has to be inside the wave packet of the particle. If it is in the lower part, the particle will end up in the lower spot. Indeed, before the wave packet reaches the Stern-Gerlach magnet, the Bohmian particle rides in a particular position of the wave packet (assuming it does not spread out significantly). When the magnet splits the wave packet to the two components moving up and down corresponding to the spin component values, the Bohmian position, being in the overlap, will continue to move horizontally (we assumed here for simplicity the constant density of the wave packets) until one of the wave packets will move away and the Bohmian position will be solely in one wave packet. If the Bohmian position was in the lower part, it will end up riding on the wave packet moving down, i.e. the spin $z$ measurement will show ``down''.

The Bohmian mechanics does not have a value for the spin. The outcome depends on the way we perform the experiment. It is considered a legitimate spin-measurement if we change the magnet in our device such that the gradient of the spin $z$ component changes its sign \cite{Ghirardi}. The only difference is that now landing in the lower spot corresponds to measuring spin $z$ ``up''. With the same initial wave function and the same Bohmian particle position, the particle, being in the lower part of the wave packet will end up in the lower spot again. But now this means that spin $z$ is ``up''. Everything is deterministic, but an observable, the spin $z$ component, does not have a definite value.

In Bohmian theory spin sometimes gets special treatment \cite{BohmSpin}, so it is better to consider the contextuality in an example which does not include spin. Let us consider position related variables $\tilde x$ and $\tilde y$ which are counterparts of the spin variables discussed in Section \ref{secEPR}, see (\ref{xtilde}) and Fig.~1.

The analogue of the Stern-Gerlach experiment above is the measurement of $\tilde y$ on a particle starting in the state $|\tilde x=1\rangle$. Similarly to the Stern-Gerlach experiment, if the Bohmian particle is placed in the right half of the initial wave packet, it will end in the right wave packet. The simple argument allowing to derive Bohmian trajectory in the Stern-Gerlach experiment is not exactly applicable here because the two wave packets (transmitted and reflected) create an interference picture when overlap, but as shown in \cite{gili}, the modification of the Bohmian trajectory can be neglected.
There is a freedom in building the beam splitter which distinguishes between the $\tilde y=1$ and $\tilde y=-1$. It can be arranged that $|\tilde y=1\rangle$ goes to the left instead of going to the right. With our initial conditions, $|\tilde x=1\rangle$ and the Bohmian particle in the right half of the wave packet, the particle, independently of the choice of the beam splitter, will be ``taken'' by the wave packet going to the right. This corresponds now to the outcome $\tilde y=-1$. Thus, we have a deterministic outcome of the experiment, but we do not have definite value of $\tilde y$. It is contextual on the design of our measuring device.

Let us consider now how Bohmian mechanics deals with the GHZ setup, Fig 2. Since each particle is maximally entangled with the two other particles, at each site there is an equal probability for every outcome of measurements of $\tilde x$ and $\tilde y$. The outcome of each measurement depends on where exactly the Bohmian position of each particle is present and what type of the beam splitter is chosen. The results of the $\tilde x_A$, $\tilde x_B$, and $\tilde x_C$, are determined by the Bohmian positions for the particles only, while the results of the $\tilde y_A$, $\tilde y_B$, and $\tilde y_C$, depend on the Bohmian positions {\it and} the types of the beam splitters. This leads to an apparent paradox: by changing the beam splitters we can flip the outcome of, say, $\tilde y_C$ measurement, but the GHZ state requires exact correlations (\ref{XYY}) between ${\tilde x}_A$, ${\tilde y}_B$, and ${\tilde y}_C$.

\begin{figure*}[h!tb]\begin{center}
 \includegraphics[width=0.75\textwidth]{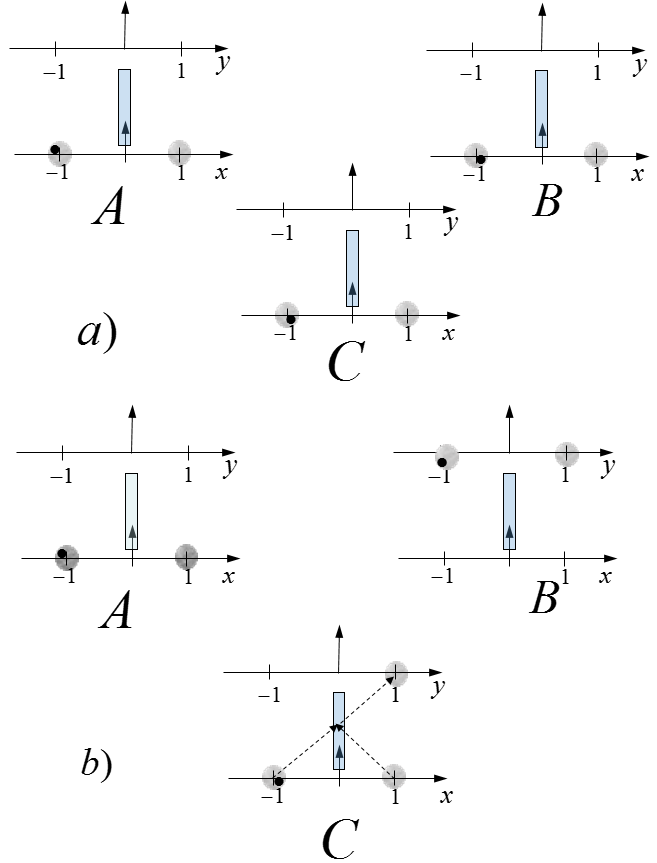}
\end{center}
\caption{a) All  Bohmian positions in the GHZ setup are placed in the wave packets $\tilde x =-1$. This ensures $\tilde x_A =\tilde x_B =\tilde x_C =-1$ and $\tilde y_A =\tilde y_B =\tilde y_C =-1$ if only one of the measurements is performed and the beam splitters are chosen with $\tilde y = 1$ on the right. b)  After the measurement of $\tilde y_B$  (passing of the wave packets through the beam splitter is enough) the conditional wave function of the particle in $C$ collapses to $|\tilde y_C=1\rangle$. This ensures  fulfillment of $\tilde x_A  \tilde y_B  \tilde y_C =1$.
 }
\label{fig:6}       
\end{figure*}

The Bohmian mechanics provides a very elegant explanation, see Fig.~6. The Bohmian positions indeed fix the outcomes of the $\tilde x$ measurements at all sites. Also, we can, by changing the type of the beam splitter, change the outcome of the $\tilde y$ measurement at each site. However, the latter is true when we perform the first $\tilde y$ measurement. After, say, $\tilde y_B$ is measured, the outcome of $\tilde y_C$ measurement becomes fixed. It should fulfill the condition $\tilde x_A~\tilde y_B~\tilde y_C=1$ and the outcome of $\tilde x_A$ is fixed even if it has not been measured yet.

The velocity formula (\ref{velo1}) explains the motion of Bohmian particles, but we can see it more easily using an equivalent, but more transparent form of this formula by separating it into
 two steps \cite{Goldstein}. First, a \textit{conditional wave function}
of particle $i$ at a particular time is defined by fixing the positions
of all other particle to be their Bohmian positions at that time:
\begin{equation}
\psi_{i}(\boldsymbol{r}_{i},t)=
\Psi\left(\boldsymbol{\textbf{r}}_{1},\ldots,\boldsymbol{r}_{i},\ldots,\boldsymbol{\textbf{r}}_{N},t\right).\label{psief}
\end{equation}
 The velocity of the $i$th particle is then:
\begin{equation}
\dot{\boldsymbol{\textbf{r}}}_{i}=
\frac{\hbar}{m}\textbf{Im}\left.\frac{\psi_{i}^{\ast}\boldsymbol{\nabla}\psi_{i}}
{\psi_{i}^{\ast}\psi_{i}}\right|_{\boldsymbol{r}_{i}=\boldsymbol{\textbf{r}}_{i}(t)}\label{rdot}.
\end{equation}

To see how the Bohmian mechanism works, let us assume that originally all Bohmian particles are in the left wave packets and that $\tilde y=1$ eigenstates move to the right.
(We cannot assume that all Bohmian particles are in the right wave packets because there is no term $|+\rangle^x_A|+\rangle^x_B|+\rangle^x_C$ in the GHZ state (\ref{GHZ1}).) This fixes the outcomes of all $\tilde x$ measurements whenever they are performed and fixes the outcome $\tilde y=-1$ for any $\tilde y$ measurement which is performed before others.
Indeed, consider the  $\tilde y_B$ measurement.
 The Bohmian position starts at the left wave packet and at the beam splitter it reaches the overlap of $|+\rangle^y_B$ and $|-\rangle^y_B$, where it stops its horizontal motion. The wave packet $|+\rangle^y_B$ continues to move and the particle is left in the wave packet $|-\rangle^y_B$ which takes it to the left. At that moment the conditional wave function at $C$ changes to $|+\rangle^y_C$. Now it does not matter that the Bohmian position at $C$ is in the left wave packet. At $C$, after the beam splitter, due to the interference of the conditional wave function, there will be only the wave packet moving to the right and the Bohmian position of particle $C$ will have to go to the right independently of its initial location. This will correspond to the outcome $\tilde y_C=1$. The requirement $\tilde x_A~\tilde y_B~\tilde y_C=(-1)(-1)1=1$ is fulfilled.

The contextuality is present here as in the case of a single particle Stern-Gerlach measurement. Changing the type of the beam splitter in $\tilde y_B$ measurement will not change the fact that the Bohmian position will go to the left in site $B$, but will change the outcome of this measurement to $\tilde y_B=1$. More importantly, the change of the beam splitter in $B$ will change the outcome of the measurement in site $C$ to $\tilde y_C=-1$. The Bohmian particle will go to the left (if the beam splitter in $C$ has not been changed). This is an action at a distance! Action at $B$ changes the outcome of the measurement in $C$ immediately after. (See Bell's discussion of this ``curious feature of Bohmian trajectories'' \cite{BellHid}.)

Another way to act on the outcome of $\tilde y_C$ measurement, which is more closely related to the Kochen-Specker theorem, is to decide not to make the measurement in $B$. Then, given our initial quantum state and the Bohmian positions in the left, the outcome will be $\tilde y_C=-1$. This will force the outcome $\tilde y_B=1$ if $\tilde y_B $ will be measured later in $B$. Although we can change outcomes by local action at a far away location, there is no way to send signals. Our preparation procedure allows us to know the quantum state of the particles, but not their Bohmian positions.

Measurement action in $B$ changes the conditional wave function in $C$. This is frequently considered as an effective collapse generated by Bohmian mechanics. This ``collapse'', however, is very different from the collapse of von Neumann (\ref{vNpro}) or the GRW collapse      (\ref{GRWpro}). The stage of a macroscopic superposition corresponding to the intermediate terms of (\ref{vNpro}) and (\ref{GRWpro}) does not appear in the Bohmian conditional wave function. Indeed, the macroscopic number of particles of the pointer will not be in a superposition of macroscopically different states for any period of time. Instead of (\ref{vNpro}) or (\ref{GRWpro}), the measurement process for the conditional wave function is a direct transition:
\begin{equation}\label{Bohmpro}
 \frac{1}{\sqrt{2}}(|{\uparrow}\rangle+|{\downarrow}\rangle)|R\rangle^{A}|R\rangle^{B}\rightarrow |{\uparrow}\rangle|Y\rangle^{A}|R\rangle^{B}.
\end{equation}
Apart from the different outcomes of the spin measurement, Fig.~5 and Fig.~4b demonstrate the difference between the ``collapse'' of the Bohmian conditional wave functions of the pointers and the GRW collapse of the wave functions of the pointers in the Stern-Gerlach experiment.

The elegant explanation of the EPR experiment shows also a serious drawback of the Bohmian theory. If a measurement in $C$ happens shortly after the measurement in $B$, then for another Lorentz observer the measurement in $C$ happens first. Then, to explain the outcome $\tilde y_C=1$, the moving Lorentz observer will need a different picture of Bohmian positions. So the elegant explanation is not covariant. Accepting the existence of a preferred Lorentz frame is a too big price for the interpretational advantages the Bohmian mechanics provides. See however some arguments for a plausibility of the preferred frame in \cite{Gold2}.

 A much less serious problem of the Bohmian interpretation is that it forces us to declare that certain position measurement devices, perfectly faithful in all other interpretations, have improper design, i.e. they incorrectly show the position of Bohmian particles \cite{gili}. The counterintuitive behavior of the Bohmian trajectories was pointed out by Bell \cite{bell2004speakable}, but the fact that such situations can lead to ``fooling'' a detector was found by Englert et al. \cite{Englert}. To be faithful, a device for the measurement of position of a particle should record it in real time in the Bohmian position of some other particle.

In my view, the most serious objection to Bohmian mechanics is that it requires the ontological status of the wave function $\Psi$ which includes the structures of multiple worlds. It does not collapse to a  wave function of a single world as von Neumann says, it does not collapse to a single  wave function of a sensible world as in the GRW theory, it remains with the MWI wave function. I do not have a good answer to the question: ``How do I know that it is not my empty wave that writes this paper and it is not your empty wave who reads it?''

We need to add a postulate that our experience supervenes on Bohmian positions and not on the wave function. This postulate does not change our mathematical theory or ontological picture of the physical Universe. But it is an important physical postulate since it tells us where {\it we} are placed in the ontology of the Universe.

In the next section I will analyze a many Bohmian worlds interpretation. It resolves the problem I raised with my question, since it avoids empty waves. It allows introducing probabilities of results of quantum experiments in a much simpler way than it is done in the pure MWI and it will help analyzing the problems of probability in other models.

\section{ Multiple Bohmian worlds}

I have not answered in a clear way two questions within a Bohmian theory formalism:   How to explain a particular choice of Nature for Bohmian positions of particles?   Why empty waves having the structure of familiar worlds do not correspond to anything in our experience? A simple suggestion avoiding both questions is to accept an actual existence of {\it all} possible Bohmian particle configurations. I am aware of the first proposal of this type by Tipler \cite{Tipler2006} (this preprint also promotes determinism of quantum theory). Tipler writes:
\begin{quote}
 ... the square of the wave function measures, not a probability density, but a density of Universes in the multiverse.

 \hfill  (Tipler, 2006)
\end{quote}
 Valentini writes about this proposal \cite{Val2010}:
\begin{quote}
 There can be no splitting or fusion of worlds. The above `de
Broglie-Bohm multiverse' then has the same kind of `trivial' structure that would be obtained if one reified all the possible trajectories for a classical test particle in an external field: the parallel worlds evolve independently, side by side. Given such a theory, on the grounds of Occam's razor alone, there would be a conclusive case for taking only one of the worlds as real.

 \hfill (Valentini, 2010)
\end{quote}
I am not convinced here. The classical and quantum cases are not identical. In classical physics there is no  preference of one trajectory relative to the other. So indeed, making all trajectories real does not help in explaining anything. In quantum case we do see a difference between worlds: we bet differently on various outcomes of quantum measurements. If density of Bohmian worlds can provide an explanation for our betting behaviour it would be a justification of Tipler's proposal.

There are other recent proposals to consider a continuum or a multitude of Bohmian worlds. They differ in what they consider ontological and some other details. All of them are trying to assign ontology to ``worlds'' in this or other form and remove the wave function from being ontological.
Bostr\"om defines \cite{Bo}:
\begin{quotation}
 A {\it world} is a collection of finitely many particles having a definite mass and a definite position. ...

A {\it metaworld} is a temporally evolving superposition of worlds of the same kind. ...

Any closed physical system is a metaworld.

... there is actually a continuum of
worlds contained in the metaworld. This continuum shall now be described by a time-dependent
universal wavefunction $\Psi_t$ in such a way that the measure
\begin{equation}\label{Bostrom}
 \mu_t(Q):=\int_Q dq |\Psi_t(q)|^2
\end{equation}
yields the {\it amount or volume of worlds} whose configuration is contained within the set $Q$...

... the wavefunction is interpreted as describing a physically existing field, and its absolute square is taken to represent the density of this field, hence a density of worlds.

 \hfill (Bostr\"om, 2012).
\end{quotation}
I will analyze below the possibility to view the wave function as a density of worlds.

Hall et al. \cite{Hall} describe the ontology of ``many interacting worlds'' (MIW). They name them ``classical worlds''. These worlds become identical to Bohmian worlds in the limit of the continuum of worlds. In their approach, there is no ontological wave function. It is a derivable concept from the evolution of worlds (through the ``reverse engineering'' of Bohmian theory). If there are a finite number of worlds, we do not reconstruct the exact wave function and the prescription is to use the wavefunction obtained by a certain averaging procedure based on the existing worlds.

The concept of interacting worlds seems to me a very artificial way to introduce interaction in physics. The theory has particles with well-defined trajectories. It is natural to introduce interaction between particles. The world can be viewed as a point in a configuration space, but known physical interactions happen in a three-dimensional space.

Sebens \cite{Se} names his paper ``Quantum Mechanics as Classical Physics''. Multiple Bohmian worlds picture he names as ``Prodigal Quantum Mechanics'' but he advocates ``Newtonian Quantum Mechanics'' which is the Prodigal Quantum Mechanics without the wave function. He writes that in Newtonian quantum mechanics the interaction is between particles, and he writes an equation (it appears as Eq.3.7) which looks like the Second Law of Newton:
\begin{equation}\label{Sebens}
 m_j \overrightarrow{a}_j =-\vec\nabla_j \left [\sum_k \frac{-\hbar^2}{2m_k}\left(\frac{\nabla^2\sqrt \rho}{\sqrt \rho}\right ) + V\right].
\end{equation}
This equation, however, includes the world density $\rho$ which makes it a very complicated dependence, more complicated than calculating $\overrightarrow{a}_j$ from the wave function.

The postulate that our experience supervenes on Bohmian particles allows simple explanation of our experience of probability. We assign probability to an event in proportion to the number of Bohmian worlds in which this event takes place. The concept of the density of worlds provides a very good explanation. Consider, for example, a particle passing through a beam splitter which reflects 30\% of the beam, Fig.~7. The equation  for the Bohmian particle velocity ensures that homogeneous distributions of a finite number of Bohmian worlds transforms into two homogeneous distributions of outgoing beams with corresponding densities.

This simple picture, however, is not applicable if there is a continuum (or even an infinite but countable number) of worlds. In this case we cannot define the concept of worlds density, see Sec. 9 of \cite{Se}. If in any region of the configuration space there is an infinity of worlds, we cannot say that it is smaller or larger than in another region.

\begin{figure}\begin{center}
  		\includegraphics[width=7.6cm]{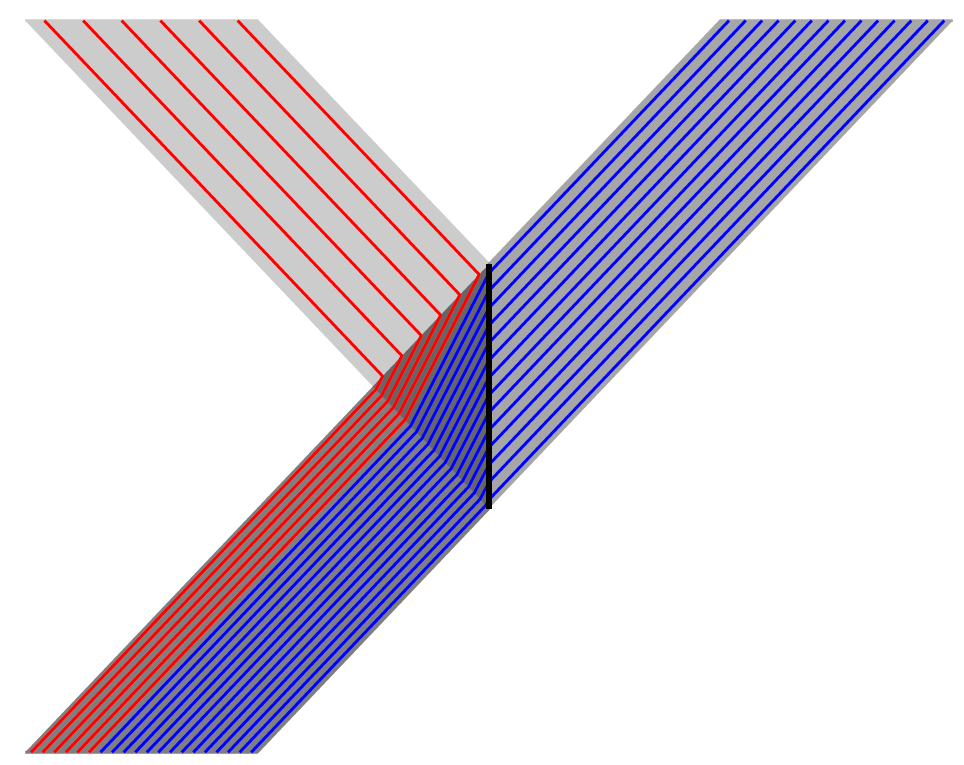}\end{center}
  \caption[Bohmian-MWI beam splitter]{A wave packet passing through a beam splitter, in the Many Bohmian Worlds picture.   In this example, there is a chance of $70\%$ to cross the beam splitter. The gray areas represent the incident, transmitted and reflected wave functions. Blue trajectories represent particles which get through the beam splitter, while red trajectories represent reflected particles. }
  \label{fig:7}
\end{figure}

The difficulty to introduce a continuum of worlds appears   in the  Bostr\"om, Hall et al., and Sebens proposals. Assigning varying measure of existence for different worlds (see Section \ref{secMeaExis}) resolves this problem, but I doubt that the authors will be ready to accept my proposal. It seems that the quotation of Tipler: ``the square of the wave function measures, not a probability density, but a density of Universes in the multiverse'' is crucial in all these approaches. Without admitting this, Bostr\"om apparently  adopts the concept of the measure of existence.  His proposal for ``volume of worlds'' (\ref{Bostrom}) includes also the ``weight'' of each world $|\Psi_t(q)|^2$. But then, I do not see how it can be considered as a density of worlds. As far as I can understand, for an infinite number of worlds there is no mathematical formalism which can provide a ``density of worlds'' picture.

I recognise the same difficulty in the influential proposal of Albert and Loewer \cite{AlLo88} for dealing with probability in the MWI by introducing infinity of ``minds''. They write:
\begin{quote}
our proposal is to associate ... an infinite set of minds in the corresponding mental state $M_K$ ... $P$ is a measure of the ``proportion'' of minds in state $M_K$.

 \hfill (Albert and Loewer, 1988)
\end{quote}
 There is no mathematical formalism which provides such a measure. Albert and Loewer wanted an infinity of ``minds'' because they wanted to ensure that for every quantum measurement, even when it will be performed in the remote future, there always be minds for all possible outcomes. Contrary to Bohmian worlds, the minds randomly move from the state ``ready'' before the measurement to the states corresponding to various outcomes. Replacing infinity of minds by a very large, but finite, number of minds   resolves the problem, but for the price of giving up the elegance and universal validity of the theory.

\section{The MWI}\label{secMWI}

 The MWI was proposed by Everett more than half century ago \cite{Eve}. For a long time it considered as a bizarre proposal, almost a science fiction. The MWI got some support from cosmologists, \cite{Tip86,AgTeg}. In quantum cosmology the MWI allows for discussion of the whole Universe, thereby avoiding the difficulty of the standard interpretation which requires an external observer. Later it was supported by the community of quantum information. It is easier to think about quantum algorithms as parallel computations performed in parallel worlds \cite{DeJo}. In recent years the MWI receives an increasing attention both in physics and philosophy journals. I want to believe that soon it will be established as the leading interpretation of quantum mechanics \cite{MWISEP}.

As I described above, the attempts to eliminate the ontology of the wave function are not really successful. We do not get a simpler ontology. Adding ontological entities beyond the wave function evolving according to the Schr\"odinger equation, like Bohmian positions or GRW hits modifying Schr\"odinger evolution are definitely useful for providing a simper connection to our experience, but they make quantum mechanics as a physical theory less attractive: the theory becomes more complicated. Moreover, the additions make dramatic conceptual changes of the theory: one adds a genuine randomness, another introduces an action at distance. So, from the point of view of a physicist, there is a tremendous advantage in avoiding additions or changes of pure quantum mechanics. The interpretational part of the theory then is much more challenging. The picture of the world we experience is not easily seen in the ontology of quantum theory. In my view, the interpretational part does have a satisfactory solution and the advantages of physics theory are by far larger than the resulting difficulties in the interpretation.

The ontology of quantum theory is the wave function of the Universe $\Psi(t)$. Nothing else. There is also the Hamiltonian of the Univerese which determines the evolution of $\Psi$, but this is a physical law, a parallel to the laws of interaction in Newtonian physics the ontology of which is positions of all particles $\vec r_i(t)$ and values of classical fields  $\vec A_j(\vec r,t)$. Here $\vec A_j$ are all classical fields: electric, magnetic, gravitational, etc.

Observables, and  their values, which are frequently considered as the basis of quantum theory, are {\it not} part of the ontology. Consequently, Heisenberg Uncertainty Relations, Robertson Uncertainty Relations, Kochen-Specker theorem, the EPR argument, the GHZ setup, and the Bell inequalities are all irrelevant for analyzing fundamental properties of Nature. They have no bearing on determinism, locality etc. I do not propose to exclude observables from quantum theory. These are useful concepts which are properties of the wave function that help to connect the wave function with our experience. The Bohmian positions do not appear in this approach to quantum theory in any status.

In classical physics there is no difficulty to add observables and their values to the ontology of the theory. The basic ontology, $\vec r_i(t)$ and $\vec A_j(\vec r,t)$, uniquely specifies the values of all observables. This, however, also tells us that adding observables to the ontology is not needed. Moreover, the connection to our experience is transparent even if only positions of particles $\vec r_i(t)$ are considered (cf Bohmian particle positions). Positions of atoms of a cat as a function of time provide a very good picture of a cat.

In the framework of the MWI it is not easy to see the connection between the wave function of atoms of a cat and our experience of a cat. The main difficulty is that  the  wave function of the Universe describes the cat in many different states and, moreover, the electrons of the cat might be in very different objects in   parallel worlds. But  it is frequently claimed that even in a hypothetical single-world Universe \cite{V-Pit} with a cat in one state, $\Psi$ cannot explain our experience \cite{Maudlin}. This is a criticism not just of the MWI, it applies also to the von Neumann Collapse and to the GRW-type collapse theories without flashes or matter density additions to ontology. The wave function $\Psi$ is defined in a high-dimensional configuration space, while our experience understood in a three-dimensional (3D) space. Formally, $N$ classical particles are also described by a point in the configuration space, but they can also be viewed as $N$ particles in 3D space. The wave function of $N$ quantum particles in the configuration space in general cannot be considered as $N$ wave functions in 3D space. If some particles are entangled we have to view the wave function in the configuration space: $\Psi(r_1, r_2)\neq\psi(r_1)\psi(r_2)$.

Yes, we do not ``experience'' entanglement. We cannot ``experience'' entanglement. ``We'' are local. Experience is a causal chain in our brain. Interactions in physics are local, so it is a chain of local and locally connected events. Surely, there is an entanglement between particles in my body. The electrons and nuclei of each atom are entangled. Atoms in every molecule are entangled. But, at the end, there is no entanglement in our experience. We do not ``feel'' the state of every electron of our body. Our experience supervenes on the state of our neurons, and, in particular, on the {\it well localized} states of our neurons. Center of mass variable of neurons are not in entangled states when we have a particular experience. In a single world of the MWI, or in the GRW world, the quantum wave packets of neurons on which our experience supervenes are well localized in the 3D space. So, we can, in principle, (we do not have yet a developed theory of our experience) view the wave function of the world as product of wave functions of relevant variables in the 3D space, multiplied by the entangled wave functions of atoms, molecules etc. responsible for the stability of matter.

We also will write an entangled state for the wave functions of qubits in a quantum computer if we have one \cite{DeJo}. The way to understand its operation is to view it as a set of parallel computations. Since the requirement for operation of a quantum  computer is a suppression of decoherence, the computations need not be considered in the 3D space.

In the MWI we introduce a concept of a ``world''. It is not an ontology: in fact, it is defined by our experience, it is a sensible story (causally connected events) that we experience. It does correspond to a set of quantum states of the Universe corresponding to that experience. Although there is a large freedom in defining what we might name as a world, I would not consider any orthogonal set of quantum states as a decomposition to worlds. My preferred definition is that in a world, by definition, all macroscopic objects are well localized \cite{MWISEP}. This is still a vague description since ``macroscopic'' and ``well localized'' are not precisely defined, but, since this concept is not about ontology of the theory, it needs not be defined with the same rigour as required in a theory about physical entities.

The locality and strength of the interactions in Nature ensure stability of worlds until we encounter a situation corresponding to a quantum measurement in which a world splits into a superposition of states each corresponding to different macroscopic descriptions. When it happens, the MWI tells that that the world splits into several worlds. This is   when von Neumann postulates that collapse makes  all but one part of the superposition to disappear. Usually, this is also when the GRW  mechanism leads to the collapse of the wave function.  Note, however,  that in the Stern-Gerlach experiment with a screen  \cite{AlV}, the splitting to stable  branches happens shortly after the atom hits the screen, while the GRW collapse mechanism might need much longer time to eliminate all but one branches.

The wave function corresponding to a world is the wave function of all particles with a property that all macroscopic objects are well localized. If we take this wave function and draw the centers of wave packets of all particles, we will obtain a picture which is very similar to the Bohmian world in which the Bohmian particle positions are somewhere inside the particle wave packets. The Bohmian world is accepted as a good description of the world, it is also very similar to the Newtonian particles world. This provides the correspondence of the wave function of a world and our experience.

The procedure which provides the correspondence between the wave function of a world and our experience is an additional physical postulate of the theory. It is not about what is in the physical Universe. The ontology is solely the wave function of the Universe. It is about {\it us}, who are {\it we} in this ontology, what corresponds to {\it our} senses. This is a counterpart of the postulate in the Bohmian theory according to which our senses supervene on Bohmian positions.

The decomposition of the wave function of the Universe into wave functions of worlds, the approximate uniqueness of the decomposition (it changes according to how fine grained we want to define our worlds) and its stability between the measurements follow from the multitude and the high density of particles around us  and the strength of electromagnetic interactions (which are responsible for most of phenomena we observe). The extensive {\it decoherence} research program, see e.g. \cite{Zur,Schlos,Zeh,Wallace} made this statement uncontroversial. This is in contrast with the early days of the MWI when the ``preferred basis'' issue considered to be a central problem of the MWI. The second main problem of the MWI, the issue of probability, which is the topic of the next two sections, remains controversial until today.

\section{The illusion of probability in the MWI}

There are two problems of probability in the MWI, one is qualitative and another is quantitative. First, it seems that the theory does not allow to define the concept of probability, the ``incoherence'' problem in the terminology of Wallace {\cite{Wa03}. Second, it is frequently claimed, and not less frequently denied, that the MWI allows a derivation of the Born rule. This section is devoted to the first problem.

Consider the Stern-Gerlach experiment, Fig.~3a when we measure the $z$ component of spin prepared in a superposition of ``up'' and ``down''. We may ask: What is the probability to get the outcome ``up''? The standard meaning is that only one of the two options might be realized: either ``up'' or ``down''. But the MWI tells that in the future there will be both. We may try to ask instead: What is the probability that ``I'' will see ``up''? The MWI tells that in the future there will be ``I'' that see ``up'' and another ``I'' that see ``down''. In the MWI I advocate, it is meaningless to ask which ``I'' shall ``I'', making the experiment, be. There is nothing in the theory which connects ``I'' before the experiment to just one of the future ``I''s.

If we are in the framework of the Many Bohmian Worlds interpretation, the problem does not arise. I, by definition, live in a world with particular Bohmian positions of all particles, and, in particular, of the particle which is measured in the stern-Gerlach experiment. The outcome is fixed before the measurement. I, however, by definition, cannot know it in advance, so we have a well-defined notion of ignorance probability.

The Albert-Loewer many minds picture \cite{AlLo88} is also not problematic. According to their construction there will be a matter of fact what outcome will see every particular mind. Since they suggest a random evolution of minds, we have genuinely random event here and a perfectly legitimate concept of probability. The  objective chance probability.

The pure MWI picture I advocate here does have a problem. There is no randomness and also there is no ignorance: all relevant information prior to the measurement is known to the observer. The question about probability of a particular outcome is illegitimate. There is no genuine concept of probability in the MWI!

But let us consider now the von Neumann collapse theory. The ontology now is a single wave function of a world, not a superposition of wave functions of worlds. The experience postulate remains the same: our experience supervenes on the wave function of the world. The wave function of the world in the single world Universe and the wave function of the world in the many-world Universe which has the same macroscopic description, are identical. Therefore, the experiences in the two worlds are also identical. In the single-world Universe with the von Neumann collapses, the concept of probability is fine: every time there is only one outcome of a quantum experiment generated by a genuinely random process.

In both theories the experiences of individual observers are the same. If the experiences are the same, they should have the same reasons to believe in one or other theory. So, it might happen that the von Neumann theory is correct, but the observer tends to believe in the MWI, or vice versa. In one theory there is a genuine probability, in another there is no probability, but the experiences are the same.

The paradox follows from the fact that we are not used to situations with multiple worlds which split. Assume that a new technology will have a machine which puts a person in and gets two identical persons out with the same memory, shape etc. You know that while you are asleep, you will be taken to this machine and the one of ``you'' will be returned home, while another ``you'' moved to another city. Moreover, you will be told all the details about the experiment, everything. It can be imagined that the complete wave function of the Universe will be given to you. In the morning, both of you, before you open your eyes, are asked: ``What is {\it your} probability to be at home?'' This is a meaningless question for everyone except for the two of you. There is a well-defined answer to this question which you, who know everything about the Universe, do not know. The question is about your identity. In which world are you? Before the experiment it was a meaningless question: you will split in two. One of ``you''  will be at home and one of ``you'' will be in another city. After the splitting, the question has a perfect sense, and if you have not opened your eyes yet, it is not a trivial question for which the probability is 0 or 1. Before opening your eyes, the two of you will have exactly the same memory state, so they should assign exactly the same probability. Both of ``you'' are related to you before the experiment, so we can formally associate the legitimate concept of probability of two of ``you'' in the morning with your illusion of probability in the evening before. It is an illusion because for having a genuine concept of probability it is required to have  only one option to be realized. In this example (as in the MWI) there is nothing which can  single out just one option.

Probability is a subtle philosophical issue. It is accepted that the de Finetti approach to it as a readiness to put an intelligent bet is as good definition as any. When you put a bet, the reward is obtained by you in the future. Since two of you in the morning have a legitimate concept of probability, they would like ``you'' in the evening to make the bet for them. This provides  the  meaning of (illusion of) probability for an observer before quantum experiment in the framework of the MWI: the ignorance probability of  his descendants \cite{schizo}. And we do not need a sophisticated technology. Ask your friends to move you while you asleep after using the iPhone ``Universe Splitter'' application ({\$}1.99) or the Tel-Aviv University World Splitter \cite{TUWS} (free). Given a laboratory with a single photon detector, you can split the world yourself. Even watching an old blinking fluorescent bulb will do, because the irregular blinking is a clearly understood quantum effect.

The resolution of the ``incoherence'' problem which I propose is still controversial. Albert \cite{Al10} claims that the probability meaning appears too late, he thinks that it is crucial to have a legitimate probability concept before performing the quantum experiment. In spite of the difficulties of attaching the trans-temporal identity to worlds, Saunders and Wallace \cite{SaWa2008a} are trying to defend the {\it diverging} instead of {\it splitting} worlds picture. Tappenden \cite{Ta2011}, however, apparently supports my position.

The ignorance meaning of probability is a subjective concept of an observer in a particular world. It is not the probability of an outcome of a quantum experiment (all outcomes take place) but the probability of  self-location of the observer after the measurement, probability that he is in the world with this outcome.
Let us turn, in the next section, to the quantitative probability problem.

\section{The measure of existence of Everett worlds}\label{secMeaExis}

Starting from Everett himself, and recently led by Deutsch and Wallace \cite{Deut,Wala}, there are attempts to prove that the counterpart of the Born rule can be derived just from the formalism of the MWI. The controversy about this subject follows, in my view, from different understandings of what is exactly assumed.

It is hard for me to take a strong side in the controversy because, on the one hand, I do see strong arguments for the Born rule, but on the other hand, I do not see a particular advantage of the MWI on this issue relative to, say, the von Neumann Collapse theory with the von Neumann cut somewhere in the measurement process.

What should be considered as ``the formalism of the MWI'' in this context? Clearly, the part of it is the statement: the ontology of our Universe is the wave function evolving according to the Schr\"odinger equation. We need also the postulate that our experience(s) supervene on the wave function only. One might say that there is no need to postulate this: we already postulated that there is nothing but the wave function, so our experience has nothing, but the wave function to supervene on. I feel that we do need to add some postulate. We need to add that our experience supervenes on the wave function in the way sketched above: we experience a cat because the wave function is roughly the product of the localized wave packets of atoms which all together have the shape of a cat. There is a logical possibility that our experience supervenes on the wave function in some other, maybe a very different way, say, through a shade the ``cat'' leaves on the wall in the Plato's cave.

As explained in the previous section, there is no real probability in the MWI, there is only an illusion of probability. So we have to derive what Tappenden named the Born-Vaidman rule \cite{schizo,Ta2011}, the rule according to which an observer should bet for his descendants, i.e. for his copies at the time after the measurement. Let me sketch now what can be viewed as a derivation of the Born-Vaidman rule (for more details see \cite{V-Pit}).

Consider a quantum experiment in which the particle wave function splits in a completely symmetrical way into three spatially separated, completely identical wave packets described by the quantum state:
\begin{equation}\label{psiin1}
 \psi ={1\over\sqrt3}(|A\rangle+|B\rangle+|C\rangle),
\end{equation}
where $|A\rangle$, $|B\rangle$ and $|C\rangle$ are the wave packets in locations $A$, $B$, and $C$. From the symmetry of the problem it follows that the probability to find the particle in $A$ is $p=\frac{1}{3}$. In \cite{V-Pit} I do it in more details ensuring that also the observer will be in three identical states. This will create three identical worlds (up to the symmetry transformations $A\rightarrow B\rightarrow C \rightarrow A $) so the probability of an observer to finds himself in any of them, and in particular in the A-world, is one third.

\begin{figure*}[h!tb]\begin{center}
 \includegraphics[width=0.75\textwidth]{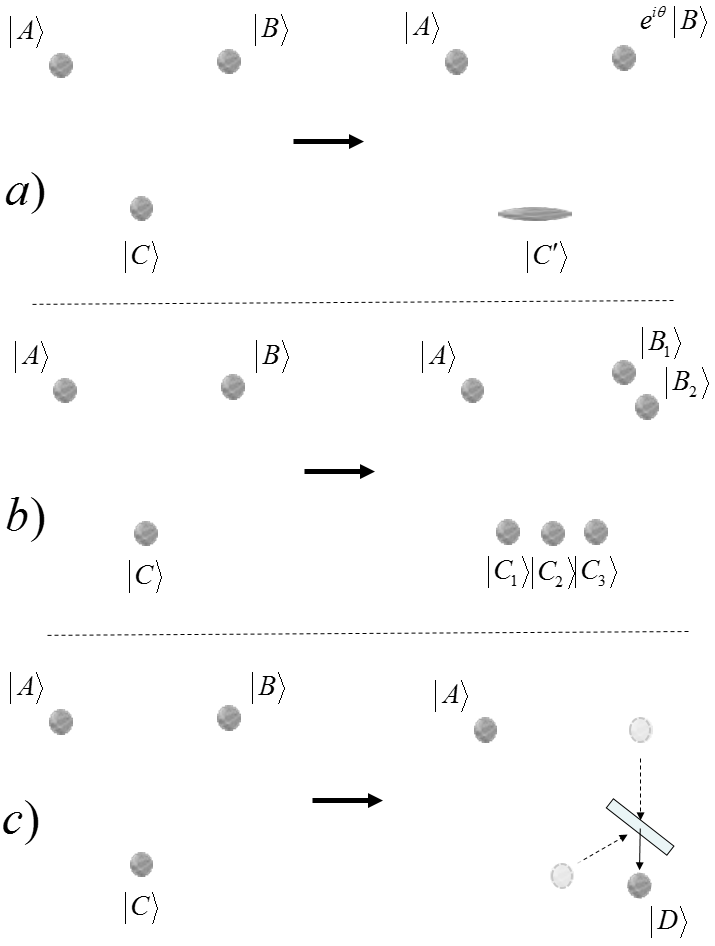}
\end{center}
\caption{Probability of finding particle in $A$ equal one third due to symmetry. Modifications of states in $B$ and $C$ cannot change the probability in $A$ since there is no action at a distance in quantum theory. a) The wave packet in $B$ gets local phase and the one in $C$ is distorted in space. b) The wave packet in $B$ splits into two identical wave packets and  the one in $C$ splits into three. c) Using a beam splitter  the wave packets in $B$ and $C$  interfere into one wave packet $|D\rangle$. }
\label{fig:8}       
\end{figure*}

 Anything which happens in a space-like separate location cannot change the probability of the outcome of an experiment in $A$, since such a change of the probability allows sending superluminal signals in direct contradiction with the special theory of  relativity. We can distort the wave packets $|B\rangle$ and $|C\rangle$ by adding a phase or by changing their shape, Fig.~8a:
\begin{equation}\label{psiin1}
 |\psi\rangle \rightarrow|\psi_1\rangle={1\over\sqrt3}(|A\rangle+e^{i\theta}|B\rangle+|C'\rangle).
\end{equation}
We can split the wave packets $|B\rangle$ and $|C\rangle$, Fig.~8b:
\begin{eqnarray}\label{psiin1}\nonumber
 |\psi\rangle \rightarrow|\psi_2\rangle={1\over\sqrt3}|A\rangle&+&{1\over\sqrt6}(|B_1\rangle+|B_2\rangle) \\
 &+&{1\over3}(|C_1\rangle+|C_2\rangle+|C_3\rangle).
\end{eqnarray}
We can make one wave packet out of the wave packets in $|B\rangle$ and $|C\rangle$ through the interference on a beam splitter, Fig.~8c:
\begin{equation}\label{psiin1}
 |\psi\rangle \rightarrow|\psi_3\rangle={1\over\sqrt3}|A\rangle+\sqrt{2\over3}|D\rangle.
\end{equation}
Moreover, the probability of finding the particle in $A$ should not be changed if we distort the local state $|A\rangle\rightarrow|A'\rangle$, otherwise it will influence the probability of detection in spatially separated location $D$.
All this tells us that the probability to find the particle in $A$ depends solely on the absolute value of the amplitude in $A$.
This is the amplitude of the wave function of the
``A-world'', the world in which the particle was found in $A$.

 I define the square of the absolute value of the amplitude of the wave function corresponding to a particular world as the {\it measure of existence} of this world. It cannot be changed unless we split the world into more worlds such that the total measure of the worlds remains the same.
When in a quantum measurement a world splits into several worlds, the observer should bet on different outcomes of the measurement in proportion to the measures of existence of corresponding worlds.

This is the Born-Vaidman rule. The rational for such betting can be as explained above: the observer does it for his future selves that do have the probability concept of self-location. The justification for betting in proportion to the measure of existence is symmetry if all the worlds have equal measures of existence. If the measures are not equal, the observer can consider a procedure in which in every world quantum measurements of some irrelevant properties are performed which split the worlds into more worlds, such that at the end of the procedure all the worlds have equal (up to a desired precision) measures of existence. Now, a natural  approach of counting worlds with a particular outcome yields the desired result \cite{Deut}.

The argument for the Born-Vaidman rule in the framework of the MWI can be transformed into the argument for the Born rule in the framework of the von Neumann Collapse approach. Until we make a measurement of where the particle is, the disturbances of the wave function are described identically in the two approaches. When we do make measurements, e.g. in the procedure of splitting worlds to the equal-size worlds, I find arguing in the framework of the MWI easier. In the collapse approach we should talk about possibilities, but the symmetry argument can be applied there too.

The advantage of the MWI approach can be seen in an elegant  resolution of the controversy in classical probability, the {\it Sleeping Beauty Problem}.
Beauty goes to sleep on Sunday. She knows that a fair coin will be tossed while she is asleep. If the coin lands Tails, then
 she will be awakened once on Monday and once on Tuesday, without having on Tuesday any memory of the previous awakening. If the
 coin lands Heads, then Beauty is awakened on Monday only. Upon each awakening, she is asked for her credence in the proposition: ``The coin has landed Heads''. Elga \cite{Elga} argued that her credence should be 1/3, while David Lewis \cite{Lewis} argued for 1/2. Since then, philosophers have been divided between halfers and thirders.

 Replacing the classical coin by a quantum coin, and analyzing the problem using the concept of the measure of existence of worlds, the credence of one third is obtained in a simple and transparent way \cite{SBAna}.
 With an obvious notation, the wave function on Monday is
\begin{equation}\label{Monday}
 |\Psi\rangle_{Mon} = {1\over\sqrt 2}(|H,~{\rm awake}\rangle_{Mon}+|T,~{\rm awake}\rangle_{Mon}).
\end{equation}
and the wavefunction on Tuesday is
\begin{equation}\label{Tue}
 |\Psi\rangle_{Tue} = {1\over\sqrt 2}(|H,~{\rm awake}\rangle_{Tue}+|T,~{\rm sleep}\rangle_{Tue}).
\end{equation}
Upon awakening, Beauty knows that she experiences one of three possible events:
either it is Monday and the coin landed Heads ($H_{Mon}$), or Monday and Tails
($T_{Mon}$), or Tuesday and Tails ($T_{Tue}$). The measures of existence of the
corresponding branches are equal, $\mu(T)=({1\over\sqrt 2})^2= {1\over  2}$. Since only one of the events
corresponds to Heads, her credence in Heads should be
\begin{equation}\label{Credence}
 Cr(H)=\frac{\mu(H_{Mon})}{\mu(H_{Mon})+\mu(T_{Mon})+\mu(T_{Tue})} =\frac{1}{3}.
\end{equation}
 Note  that Peter
Lewis \cite{PeLew} also addressed the Sleeping Beauty problem  in the MWI framework, but advocated the halfer solution. The concept of the measure of existence allows a transparent analysis of his error, see \cite{SBAna}.

In the discussion above I have shown a manifestation of the measures of existence of future worlds which will be created after the measurement. Is the measure of existence of the present world has a physical manifestation? An observer does not feel the measure of existence of the world she lives in. There is no experiment she can perform that  distinguishes between large and small measures of existence. Nevertheless, I do see a difference \cite{schizo}. There is a hypothetical situation in which she should behave differently based on her knowledge of the measure of existence of her present world. She might know the relative measure of existence of her world and a parallel world if she  performed  a quantum experiment a minute ago. The hypothetical situation which manifests the difference includes  an alien with a super technology  which convinces the observer that he can make interference experiments with macroscopic objects. The alien offers a bet to the observer about the outcome of another experiment that she will perform in her laboratory. Now, the measure of existence of the present world matters. The observer suspects that the alien will use a parallel world to interfere and change the odds of her quantum experiment. If the measure of existence of her world is smaller than that of the parallel world, then she should refuse to make any bet since alien can arrange any outcome with certainty. If however, her world has larger measure of existence than the parallel world, the alien can change the probability only up to some limit, so she can place some bets.

We can be in the role of the alien for a photon in a laboratory. A photon reaching a second  $50\%/50\%$ beam splitter in a Mach-Zehnder  interferometer ``thinks'' that it can pass the beam splitter and reach a detector behind it  with probability half. We, however can tune the interferometer  in such a way that it will be a dark port, so the photon has zero chance to reach the detector.

The name ``measure of existence'' suggests ontology, but ``worlds'' are not part of the ontology of the MWI that I advocate. ``I'', people are also not part of the ontology. We are patterns,  shapes  of the wave function.  I and my world certainly exist and are ``real'', but not in the sense of the reality of the wave function. It might be useful to introduce some new semantics, something like ``physical reality'' and ``reality of experience'', ``physical ontology'' and ``experience ontology''. Note that the term ``primitive ontology'' has already been introduced  \cite{AllGT}. It is not the wave function of the Universe (as it lives in the 3D space) and it is not defined by experience. Too many concepts of ontology might lead to a confusion.

 \section{(Non)Locality}

When we consider the physical Universe and do not think about us in this Universe, we have a theory which describes it in a deterministic and, in some sense, local way. The wave function of the Universe $\Psi$ at one time determines completely $\Psi$ at all times. Without collapse in quantum measurement the EPR-Bell-GHZ nonlocality is not present in quantum theory: local actions change nothing in remote locations.

Interactions are local in 3D space, so a simple and coherent picture would be waves of all particles locally interacting with each other. The wave function of the Universe is not like this due to entanglement, but it can be decomposed into a superposition of products of wave functions of all particles. Decomposition into a superposition of products of wave functions of molecules or slightly bigger objects corresponds to  the decomposition into the superposition of  wave functions of  worlds of the MWI.

In classical theory, there is a simple local picture of interactions. Particles create fields in a local way. These fields, or change in these fields spread out in space not faster than light. Then particles feel local forces due to these fields and change their velocities. In quantum theory, the interactions also local, but this Newtonian picture does not exist. There is (a complicate)  counterpart of the Second Law of Newton  for Bohmian particles (\ref{Sebens}), but not for the wave function.

The Schr\"odinger and other equations for the evolution of the wave function are based on potentials, not on local fields. This picture is not explicitly local since potentials do not have definite local values: they can be changed  through gauge transformations without any physical change. Only some global properties of potentials (such as integrals on closed curves) are gauge invariant. One of such integrals, the line integral of the vector potential of the electromagnetic field, plays a central role in the Aharonov-Bohm (AB) effect \cite{AB}. It can be observed using an interference experiment with a charged particle even if the particle never passes through a region with an electromagnetic field.

The AB effect convinced the physics community that in quantum theory (in contrast to classical theory) potentials are not just auxiliary constructions for calculating fields, but have direct physical effects. It means that a simple classical picture with particles creating fields which locally affect motion of other particle cannot be true in quantum theory. Quantum theory cannot be local in the form of locality of classical physics. (The ``Second Law of Newton'' for Bohmian particles (\ref{Sebens}) can be considered local, but the Bohmian theory is manifestly nonlocal due to the possibility of remote change of $\rho$.)

My hope is that the MWI removes the nonlocality from quantum theory. By denying the existence of the collapse of the wave function, the MWI removes the action at a distance due to quantum measurements, but  the absence of collapse does not help removing nonlocality of the AB effect. Although the AB effect does not provide an action at a distance,  it apparently does not allow  a local explanation of the evolution of the quantum wave function.

Recently, however, I proposed a local explanation of the AB effect \cite{VAB}. In the standard approach to the AB effect, the electromagnetic field and its source are classical. But classical physics was shown to be incorrect experimentally, so for the precise analysis everything should be considered quantum. Building a model of a solenoid in the AB setup as two oppositely charged cylinders rotating in opposite directions I could consider the source of the magnetic field in the framework of the quantum theory. I have shown that the field of the electron passing through two sides of the solenoid causes, due to the local forces on the cylinders, small rotations in opposite directions. Considering the cylinders as quantum objects, the tiny relative rotation transforms into the relative phase in the wave functions of the cylinders. In the beginning of the AB experiment, the electron becomes entangled with the cylinders. In the end of the process, the electron and the cylinders are again in the product state, but the relative phase acquired by the cylinders is transformed to the electron which exhibits the AB effect.  I have shown that the relative phase in the cylinders is exactly equal to the AB phase, thus providing a local explanation of the AB effect.

This result leaves me a hope that one day there will be a local version of quantum mechanics, the counterpart of Newton's Laws formulation of classical physics. Even for classical physics, global approaches starting with Hamiltonian or Lagrangian are more efficient and can explain everything. And I do not foresee that in quantum theory there will be some effects not explainable in a global way. But, that does not necessarily mean that there might not be an alternative, equally valuable, local description.

The existence (or the nonexistence) of the local field picture is a very important feature of Nature. It has  testable consequences even before such picture is discovered. If such theory exists, it follows that when all particles move in field free regions, no effect of these fields and its potentials can be observed.  A particular version of the Electric AB effect corresponds to such situation  \cite{VAB}. In this setup  there are potentials but the electromagnetic fields vanish at locations of every particle. Careful calculations \cite{Tirzamod} based on electromagnetic potentials which take all systems into account show that, indeed, in this setup the AB effect vanishes.

The physical Universe is local in the sense that there is no action at a distance. In a particular gauge, local changes of the wave function are explained by local values of potentials. There is also a hope that one might find a gauge independent formalism with local actions. But there are connections between different parts of the Universe, the wave function of the Universe is entangled. Entanglement is  the essence of the nonlocality of the Universe. ``Worlds'' correspond to sets of well localized objects all over in space, so, in this sense, worlds are  nonlocal entities. Quantum measurements performed on entangled particles lead to splitting of worlds with different local descriptions. Frequently such measurements lead to quantum paradoxes which will be discussed in the next section.

\section{Resolution of Quantum paradoxes in the framework of the MWI}

Paradoxical nonlocal phenomena in a single-world picture obtain local explanation in the framework of the MWI.
The nonlocality of the GHZ and other EPR-Bell type situations follows from the nonlocality of ``worlds''. By definition, macroscopic objects are well localized in each world, but at least some of the objects are localized in a different way in various worlds. Spatially separated entangled particles, through local interaction with macroscopic objects, create worlds with nonlocal correlations. Measurement of one particle of the EPR pair changes nothing for the other particle if it is considered in the Universe, but it creates worlds with definite spin of a remote particle.

If the EPR pair is used for teleportation, the Bell measurement in one site creates four worlds with the quantum state teleported to the second particle of the pair and rotated in  four definite ways. The mixture of these four states corresponds to a completely unpolarized density matrix, the description of the particle of an undisturbed  EPR pair. Thus again, from the point of view of the Universe, no change in  the  second EPR particle took place.

Another paradoxical example is an interaction-free measurement (IFM) \cite{EV}. We can sometimes (and in an improved setup \cite{IFMzeno} almost always) find an opaque object, i.e. to know with certainty that it is present in a particular place, without visiting this place. We get information about the place without any particle passing through, without a particle reflecting from it and even without a particle being near this place.
The operational meaning that a particle was not in a particular place is that it left no trace of any strength at this place.

 The basic IFM is just a Mach-Zehnder interferometer tuned to a complete destructive interference in one of the output ports running with single photons. We test that there is nothing everywhere inside the interferometer except for a one place on the path of one of the arms of the interferometer. If a single run provides a click in the dark port, we know that in this place there is an object. On the other hand, we are certain that the photon left zero trace in the arm of the interferometer where the object was present since a single photon cannot be found in two places.

In the one world picture this is a very paradoxical situation: we cannot see where is the causal link between the presence of an object and our knowledge about the presence of the object. But physics describes all worlds together. In a parallel world the particle was absorbed by the object. It left a (weak) trace near the object.

Based on the IFM, similar paradoxical tasks have been suggested and achieved.
Counterfactual computation \cite{Jozsa}, in which we obtain the result of the computation without running the computer, has the same resolution in the MWI framework: in parallel worlds the computer has been running. Counterfactual cryptography \cite{Noh,exp} based on the fact that the particle carrying the information was not present in the area where Eve could measure it. In parallel worlds the particle was near Eve, but the cryptographic protocol was cleverly arranged such that in these parallel worlds  the transmitted key has been aborted.

It has also been claimed that ``direct counterfactual communication'' \cite{Zub,Cao}, ``counterfactual entanglement distribution'' \cite{Guo}, and ``counterfactual transporting of an unknown qubit'' \cite{Salih}, are possible. In these protocols, indeed, the particle which transmits the information could not pass through the transmission channel. However, I find that these protocols cannot be named ``counterfactual'' \cite{VcomSalih}. The particle cannot pass through the transmission channel, but it leaves a trace there. Recent experiment with a similar setup demonstrates this trace \cite{Danan}. So, my conclusion is that we cannot say that the particle was not there. (Note that the authors of ``direct counterfactual communication protocol'' were not convinced \cite{RVCom,ComPast,RComPast}.)

\begin{figure}\begin{center}
 	  \includegraphics[width=7.6cm]{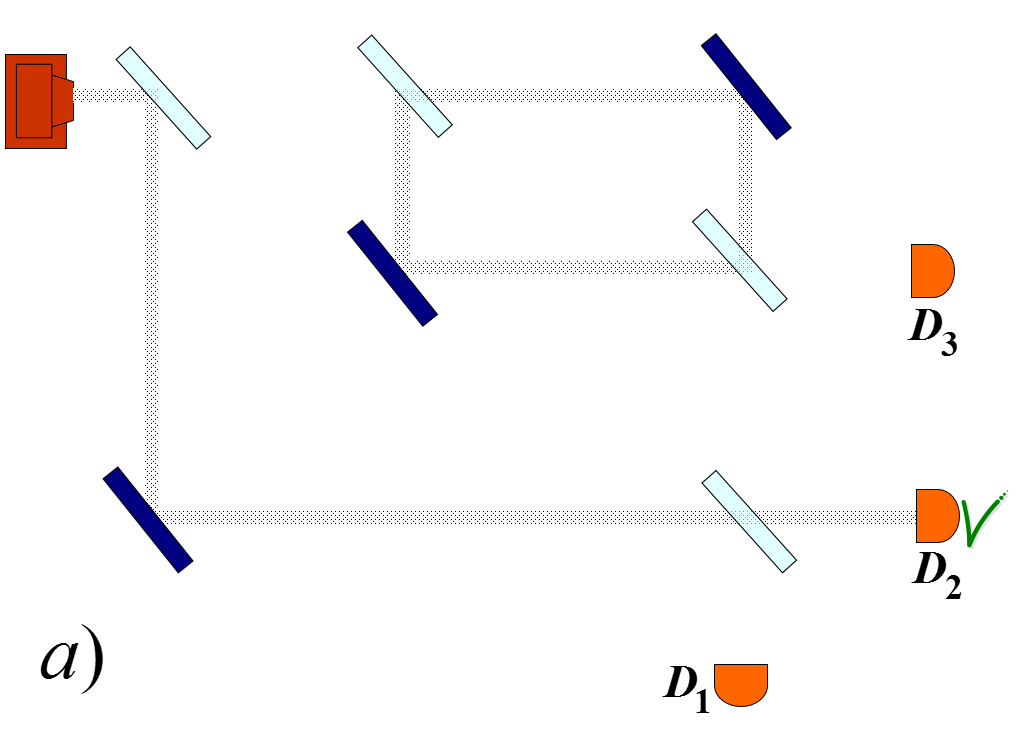}\\
 \includegraphics[width=7.6cm]{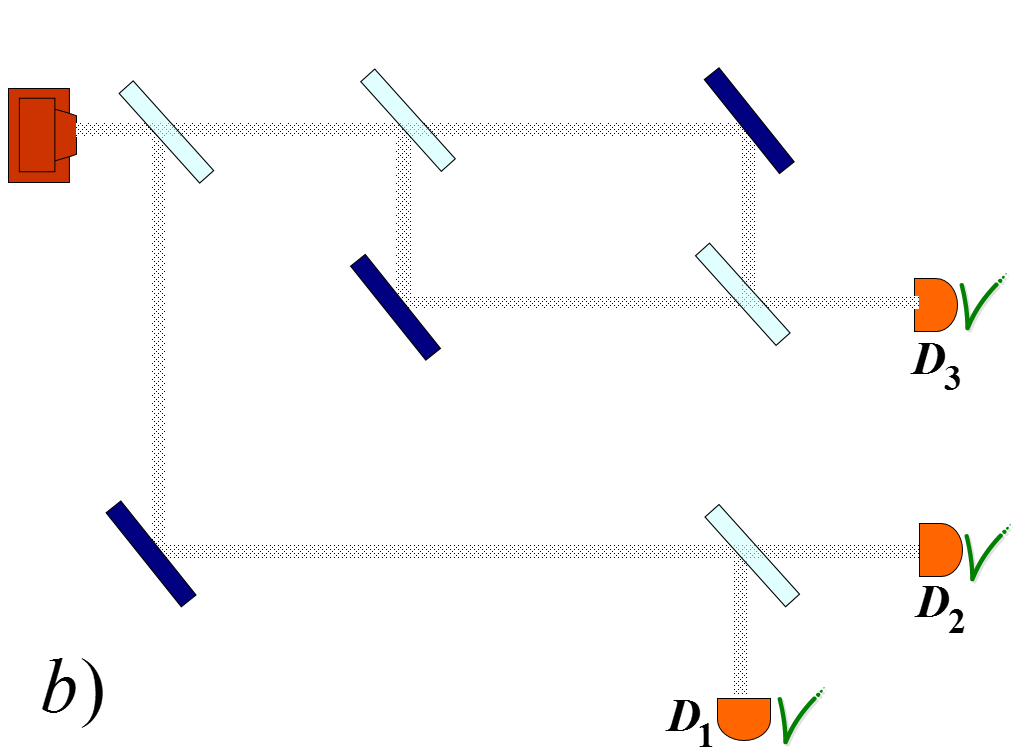}\end{center}
  \caption[weak trace]{(a). The weak trace left by the photon entering the interferometer in the world in which it was detected in dtector $D_2$. (b). The weak trace left by the photon in the physical Universe. }
  \label{fig:9}
\end{figure}

I use the criterion that the particle was where it left a weak trace, so the particle was in the transmission channel. The paradoxical feature of the particle's trace is that the trace is not continuous from the source to the detector. In the example conceptually similar to all these protocols, the trace appears inside an inner interferometer, but there is no trace which leads towards (and outside) the interferometer \cite{Vpast}, see Fig.~9a. Again, this paradoxical situation   happens   in a single world in which the photon was detected by $D_2$. In the physical Universe, and in the three parallel worlds in which the photon entered the interferometer and was detected by detectors $D_1$, $D_2$, and $D_3$ taken together,  the  trace which photon leaves is not paradoxical, see Fig.~9b.

What I mean by ``not paradoxical trace'' is that it is continuous, we do not have a separate island of trace as in Fig.~9a. But there is still another paradoxical feature: one particle leaves simultaneously trace in a few separate places. Although the trace is weak, there are experiments which can identify the local trace with certainty (these experiments succeed only rarely). The weak trace in Fig.~9b is entangled, so in a world in which strong trace is discovered in one of the paths, the traces in other locations  disappear immediately.  Considering a history of a world with such measurements, we see actions at a distance due to these measurements. There is no action at a distance only in the physical Universe, where all the worlds are considered together.

\section{Conlusions}

In this paper I reviewed the interpretations and the fundamental aspects of quantum mechanics, arguing that, contrary to a popular view, quantum theory can be considered as a deterministic theory describing Nature.

The theory has two parts. The first part is physical, mathematical, the one to which I attach the words ``ontological'' or ``ontic'' (without attributing a distinction between them).  It is a counterpart of the theory of particles and fields in classical physics. This  is the part of a  theory about what {\it is} in the physical Universe. Although I discussed several approaches, I find by far the best option   to take the wave function of the Universe, and only it, as the ontology of the theory. Major part of the paper explains why I have this view. I also review numerous recent  works on the subject  pointing in this direction.

The theory of the wave function  is a deterministic theory without action at a distance.  It is the theory about what is, irrespectively of us. Even quantum observables, like momentum, energy, spin, etc. which are frequently considered to be the starting point of quantum mechanics, are not considered ontological in this approach. Thus, various uncertainty relations between quantum variables do not lead to indeterminism.

The second part of the theory makes the connection to us and our experience. I do not use the words ``ontological'' and ``ontic'' for that part: experiences,  people, chairs, and even our worlds are not ontological in this view. Our experiences supervene  on the wave function of the Universe but in a non trivial way. Since observables are not ontological, connection between experience and ontology requires elaborate construction. The main reason for the difficulty is that the ontology, the wave function of the Universe, corresponds to multiple experiences. Thus, we need to define the concept of a ``world'', the concept of ``I'', etc. All these are just properties, shapes of the wave function, and in fact, the shape of a particular  term in the  superposition which constitutes the wave function of the Universe. Although the detailed correspondence is difficult, the locality and strength of physical interactions suggest that such a program is feasible and for small systems for which it has been implemented, it never led to contradictions.

I hope that this work will trigger further analysis: extending this picture to quantum field theory,  tightening the gaps in understanding of our senses,  clarifying the philosophical concepts. Science needs to reach a consensus regarding interpretation of quantum mechanics and I feel that physics tells us that this is the most promising direction.

The theory of Universal wave function is deterministic, local, free of paradoxes,  and fully consistent with our experience.  I do not see  ``clouds'' in the beauty and clearness of quantum theory similar to ``two clouds''  Lord Kelvin saw in 1900 in classical physics.

I thank Kim  Bostr\"om, Eliahu Cohen, Roger Colbeck, Bob Doyle, Naama Hallakoun, Carl Hofer, Yitzhak Melamed,  Zeev Schuss, and Charles Sebens for helpful discussions. This work has been supported in part by grant number 32/08 of the Binational Science Foundation and the Israel Science Foundation Grant No. 1125/10.


\begin{thebibliography}{999}

\bibitem{Piron}
S. Pironio, A. Acin, S. Massar, A. Boyer de la Giroday, D. N. Matsukevich, P. Maunz, S. Olmschenk, D. Hayes, L. Luo, T. A. Manning, and C. Monroe,
Random Numbers Certified by Bell's Theorem,
Nature 464, 1021 (2010).


\bibitem{MWISEP} L.~Vaidman,  Many-Worlds Interpretation of Quantum
Mechanics, {\it Stan. Enc. Phil.},  E. N. Zalta (ed.) (2014),
http://plato.stanford.edu/entries/qm-manyworlds/.

\bibitem{Lapl1}
P. S. de Laplace
Essai Philosophique sur les Probabilites [1814], in
{\it Academy des Sciences, Oeuvres Complotes  de Laplace}, Vol. 7, Gauthier-Villars, Paris (1886).

\bibitem{Leuc}
Leucippus, Fragment 569 - from Fr. 2 Actius I, 25, 4, in
   {\it The Atomists: Leucippus and Democritus. Fragments, A Text
and Translation with Commentary}, C.C.W. Taylor, Toronto: University of Toronto Press (1999).


\bibitem{Spin}
 B. Spinoza,
  {\it Ethics},
   Proposition 29, Part 1,  in
  {\it The Collected Writings of Spinoza}, E. Curley, trans., Princeton University Press, Vol. 1, Princeton (1985).


\bibitem{Leib}
G. Leibniz, in H. G. Alexander, ed.,
{\it The Leibniz-Clarke correspondence},
 New York: Barnes and Noble  (1956).

\bibitem{Russ}
B. Russell, {\it Our knowledge of the external world},
New York: Menton Books (1956).

\bibitem{Earm}
J. Earman,
{\it A primer on determinism},
 Western Ontario Series in Philosophy of Science, Vol. 32, D. Reidel (1986).

\bibitem{Lapl2}
P.S. de Laplace,
{\it Theorie Analytique des Probabilites,}
 Courcier, Paris (1820).

\bibitem{deFin}
B. de Finetti, {\it Theory of Probability,} Vol. 1. New York: John Wiley and
Sons (1974).

\bibitem{Accardi}
L. Accardi,
Topics in Quantum Probability,
  Phys. Rep.  77, 169  (1981).

\bibitem{Frohl}
J. Fr\"ohlich, B. Schubnel,
Quantum probability theory and the foundations of quantum mechanics,
arXiv:1310.1484 (2013).

\bibitem{Spek07}
R.W. Spekkens,
Evidence for the epistemic view of quantum states: A toy theory,
Phys. Rev. A  75, 032110 (2007).

\bibitem{BivonNE}
G. Birkhoff,  and J. von Neumann,
The logic of quantum mechanics,
Ann.   Math. 37, 823 (1936).

\bibitem{Pit}
I. Pitowsky
Quantum mechanics as a theory of probability,
in
{\it Physical Theory and its Interpretation,}
The Western Ontario Series in Philosophy of Science,  Vol 72,  213  (2006).

\bibitem{CFS}
C.M. Caves, C.A. Fuchs, and R. Schack,
Quantum probabilities as Bayesian probabilities,
Phys. Rev. A 65,   022305   (2002).

\bibitem{Qbist2}
 C.A. Fuchs, and R. Schack,
 Quantum-Bayesian coherence,
Rev. Mod. Phys. 85, 1693 (2013).

\bibitem{PR}
S. Popescu and D. Rohrlich,
Quantum nonlocality as an axiom,
 Found. Phys. 24, 379 (1994).

\bibitem{Ha5Ax}
L. Hardy,
Probability theories in general and quantum theory in particular,
 Stud. Hist. Phil. Mod. Phys. 34B, 381 (2003).

\bibitem{DaLe}
E. B. Davies and J. T. Lewis,
An operational approach to quantum probability,
Commun. Math. Phys. 17, 239 (1970).

\bibitem{JaHi}
P. Janotta and H. Hinrichsen,
Generalized probability theories: What determines the structure of quantum physics?
arXiv:1402.6562, (2014).


\bibitem{Grif84}
R. B. Griffiths,
 Consistent histories and the interpretation of quantum mechanics,
 J. Stat. Phys. 36, 219 (1984).


\bibitem{GellHart1990}
M. Gell-Mann and J. B. Hartle,
 in {\it Complexity, Entropy and the Physics of Information},  W. H. Zurek, ed. (Addison-Wesley, Redwood City, CA, ), p. 425 (1990).

 \bibitem{Grif2013}
 R. B. Griffiths,
 A consistent quantum ontology
   Stud. Hist. Philos. Mod. Phys. 44, 93 (2013).


\bibitem{BaWi}
H. Barnum, A. Wilce,
Post-classical probability theory,
in {\it Quantum Theory: Informational Foundations and Foils}, G. Chiribella and R.W. Spekkens, eds.,  Springer, forthcoming.
	arXiv:1205.3833.

\bibitem{CoeSp}
B. Coecke and R. W. Spekkens,
Picturing classical and quantum Bayesian inference,
Synthese 186,  651 (2012).

\bibitem{LeiSp}
M. S. Leifer and R. W. Spekkens,
Towards a formulation of quantum theory as a causally neutral theory of Bayesian inference,
Phys. Rev. A 88, 052130 (2013).

\bibitem{Hei}
W. Heisenberg,
ÜUber den anschaulichen Inhalt der quantentheoretischen Kinematik und Mechanik,
Z. Phys. 43, 172 (1927).

\bibitem{Oza2003}
M. Ozawa,
Physical content of Heisenberg's uncertainty relation: limitation and reformulation,
Phys. Lett. A, 318, 21 (2003).

\bibitem{Erhard}
J. Erhart, S. Sponar, G. Sulyok, G. Badurek, M. Ozawa, and Y. Hasegawa,
Experimental demonstration of a universally valid error-disturbance uncertainty relation in spin measurements,
Nature Phys. 8, 185 (2012).


\bibitem{Rozema}
L.A. Rozema, A. Darabi, D.H. Mahler, A. Hayat, Y. Soudagar, and A.M. Steinberg,
Violation of Heisenberg's measurement-disturbance relationship by weak measurements,
Phys. Rev. Lett. 109, 100404 (2012).

\bibitem{Bush}
P. Busch, P. Lahti, and R. F. Werner,
Proof of Heisenberg's error-disturbance relation,
Phys. Rev. Lett. 111, 160405 (2013).


\bibitem{Dressel-Nori}
J. Dressel and F. Nori,
Certainty in Heisenberg's uncertainty principle: Revisiting definitions for estimation errors and disturbance,
Phys. Rev. A  89, 022106 (2014).

\bibitem{Deu83}
D. Deutsch,
Uncertainty in quantum measurements,
Phys. Rev. Lett. 50, 631 (1983).

\bibitem{Friedland}
S. Friedland, V. Gheorghiu, and G. Gour,
Universal uncertainty relations,
Phys. Rev. Lett. 111, 230401 (2013).


\bibitem{Rob}
H. P. Robertson,
The uncertainty principle,
Phys. Rev. 34, 163 (1929).



\bibitem{Vdadb}
 L. Goldenberg and L. Vaidman,
Applications of a simple quantum mechanical formula,
Am. J. Phys. 64, 1059 (1996).

\bibitem{KS}
S. Kochen, and  E. Specker,
The problem of hidden variables in quantum mechanics,
Jour. of Math. Mech.  17, 59 (1967).

\bibitem{Peres}
A. Peres,
Two Simple proofs of the Kochen-Specker Theorem,
J. Phys. A 24,   L175 (1991).

\bibitem{KS18}
A. Cabello, J.M. Estebaranz, and G. Garc\'{\i}a-Alcaine,
Bell-Kochen-Specker Theorem: A proof with 18 vectors,
 Phys. Lett. A  212,   183 (1996).

\bibitem{KS13}
S. Yu and C. H. Oh,
State-independent proof of Kochen-Specker Theorem with 13 rays,
Phys. Rev. Lett. 108, 030402 (2012).

\bibitem{Abbot}
A.A. Abbott, C.S. Calude, J. Conder, and K. Svozil,
Experimental test of quantum contextuality in neutron interferometry,
Phys. Rev. A 86, 062109 (2012).

\bibitem{Me}
D.A. Meyer,
Finite precision measurement nullifies the Kochen-Specker theorem,
Phys. Rev. Lett. 83, 3751 (1999).


\bibitem{Kent}
 A. Kent,
Noncontextual hidden variables and physical measurements,
Phys. Rev. Lett. 83, 3755 (1999).

\bibitem{Ca08}
 A. Cabello,
Experimentally testable state-independent quantum contextuality,
Phys. Rev. Lett. 101, 210401 (2008).


\bibitem{Cab1}
G. Kirchmair, F. Z\"ahringer, R. Gerritsma, M. Kleinmann, O. G\"uhne, A. Cabello, R. Blatt and C. F. Roos,
State-independent experimental test of quantum contextuality,
Nature 460,  494-497 (2009).

\bibitem{Cab2}
E. Amselem, M. Radmark, M. Bourennane and A. Cabello,
State-independent quantum contextuality with single photons,
Phys. Rev. Lett. 103,  160405 (2009).



\bibitem{CSneutr}
H. Bartosik, J. Klepp, C. Schmitzer, S. Sponar, A. Cabello, H. Rauch, and Y. Hasegawa,
Experimental test of quantum contextuality in neutron interferometry,
Phys. Rev. Lett. 103, 040403 (2009).


\bibitem{KSphott}
V. D'Ambrosio, I. Herbauts, E. Amselem, E. Nagali, M. Bourennane, F. Sciarrino, and A. Cabello,
Experimental implementation of a Kochen-Specker set of quantum tests,
Phys. Rev. X  3, 011012 (2013).

\bibitem{EPR}
A. Einstein, B. Podolsky, and N. Rosen,
Can quantum-mechanical description of physical reality be considered complete?
Phys. Rev. 47, 777 (1935).


\bibitem{BA}
D. Bohm and Y. Aharonov,
Discussion of experimental proof for the paradox of Einstein, Rosen, and Podolsky,
Phys. Rev. 108, 1070  (1957).

\bibitem{Bell1}
J. S. Bell,
On the Einstein-Podolsky-Rosen Paradox,
 Physics 1, 195 (1964).

\bibitem{GHZ}
D. M. Greenberger, M. A. Horne, and A. Zeilinger,
Going beyond Bell's theorem,
 in {\it Bell Theorem , Quantum Theory and Conceptions of the Universe}, M. Kafatos, ed., p. 69, Kluwer Academic, Dordrecht (1989).

\bibitem{Mer}
4. N. D. Mermin,
 Quantum mysteries revisited,
 Am. J. Phys. 58, 731 (1990).

\bibitem{VGHZ}
L. Vaidman,
Variations on the theme of the Greenberger-Horne-Zeilinger proof,
Found. Phys. 29, 615  (1999).

\bibitem{Ballentine}
L.E. Ballentine,
 The statistical interpretation of quantum mechanics,
 Rev. Mod. Phys. 42, 358 (1970).

\bibitem{Nelson}
E. Nelson,
Derivation of the Schr\"odinger Equation from Newtonian mechanics,
Phys. Rev. 150, 1079 (1966).

\bibitem{PBR}
M. F. Pusey, J. Barrett, and T. Rudolph,
On the reality of the quantum state,
Nature Phys. 8, 476 (2012).

\bibitem{Hardy}
L. Hardy,
Are quantum states real?
Int. J. Mod. Phys. B   27, 1345012   (2013).

\bibitem{PaMas1}
M. K. Patra, S. Pironio, and S. Massar,
No-go theorems for $\psi$-epistemic models based on a continuity assumption,
Phys. Rev. Lett. 111, 090402 (2013).


\bibitem{Patraexpereiment}
M. K. Patra, L. Olislager, F. Duport, J. Safioui, S. Pironio, and S. Massar,
Experimental refutation of a class of $\psi$-epistemic models,
Phys. Rev. A 88, 032112 (2013).


\bibitem{CR2}
R. Colbeck and R. Renner,
Is a system's wave function in one-to-one correspondence with its elements of reality?
Phys. Rev. Lett. 108, 150402 (2012).

\bibitem{CR1}
R. Colbeck and R. Renner,
No extension of quantum theory can have improved predictive power,
Nature Commun. 2, 411 (2011).

\bibitem{GRFound}
G.C. Ghirardi, and R. Romano,
Ontological models predictively inequivalent to quantum theory,
Phys. Rev. Lett. 110, 170404 (2013).


\bibitem{Lad}
F. Laudisa,
Against the `no-go' philosophy of quantum mechanics,
Euro. Jnl. Phil. Sci. 4, 1 (2014).

\bibitem{Bell-no-go}
J.S. Bell,
On the impossible pilot wave,
Found. Phys. 12, 989 (1982).


\bibitem{CoKo}
J. Conway and S. Kochen,
The Free Will Theorem,
Found. Phys. 36, 1441 (2006).



\bibitem{HaSp}
N. Harrigan and R. W. Spekkens,
Einstein, incompleteness, and the epistemic view of quantum states,
Found. Phys. 40, 125 (2010).

\bibitem{Montana}
A. Montina,
State-space dimensionality in short-memory hidden-variable theories,
Phys. Rev. A 83, 032107 (2011).

\bibitem{Leifer}
M. S. Leifer,
$\psi$-epistemic models are exponentially bad at explaining the distinguishability of quantum states,
 arXiv:1401.7996 (2014).

\bibitem{GaHa}
E. Galvao and L. Hardy,
Substituting a qubit for an arbitrarily large number of classical bits,
Phys. Rev. Lett. 90, 087902 (2003).


\bibitem{VaMit}
L. Vaidman and Z. Mitrani,
Qubits versus bits for measuring an integral of a classical field,
Phys. Rev. Lett. 92, 217902 (2004).


\bibitem{PMeas}
Y. Aharonov and L. Vaidman,
 Measurement of the Schr\"odinger wave of a single particle,
Phys. Lett. A 178, 38  (1993).

\bibitem{Got}
K. Gottfried,
 Does quantum mechanics describe the ``collapse'' of the wave function?,
  contribution to the Erice school: {\it 62 Years of Uncertainty} (unpublished) (1989).


\bibitem{Psi2}
A. Montina,
Epistemic view of quantum states and communication complexity of quantum channels,
Phys. Rev. Lett. 109, 110501 (2012).

\bibitem{Psi3}
M. Schlosshauer and A. Fine,
Implications of the Pusey-Barrett-Rudolph quantum no-go theorem,
Phys. Rev. Lett. 108, 260404 (2012).

\bibitem{Psi4}
M. Leifer and O. Maroney,
Maximally epistemic interpretations of the quantum state and contextuality,
Phys. Rev. Lett. 110, (2013).

\bibitem{Psi5}
S. Aaronson, A. Bouland, L. Chua, and G. Lowther,
$\psi$-epistemic theories: The role of symmetry,
Phys. Rev. A 88, 032111 (2013).


\bibitem{Psi6}
M. Leifer,
Is the quantum state real? A review of $\psi$-ontology theorems,
Quanta, to be published, http://mattleifer.info/wordpress/wp-content/uploads/2008/10/quanta-pbr.pdf


\bibitem{Psi7}
M. Schlosshauer and A. Fine,
No-go theorem for the composition of quantum systems,
Phys. Rev. Lett. 112, 070407 (2014).


\bibitem{Lewis1}
P. G. Lewis, D. Jennings, J. Barrett, and T. Rudolph,
Distinct quantum states can be compatible with a single state of reality,
Phys. Rev. Lett. 109, 150404 (2012).



\bibitem{vonNeuma}
J. von Neumann,
 {\it Mathematical foundations of quantum theory and measurement},
 Princeton University Press, Princeton, (1983).

\bibitem{againstmeasurBell}
J. S. Bell
Against ``measurement'',
in {\it Sixty-Two Years of Uncertainty}, NATO ASI Series Vol.  226, p. 17 (1990).

\bibitem{Pear1}
P. Pearle,
Reduction of the state vector by a nonlinear Schr\"odinger equation,
Phys. Rev. D 13, 857 (1976).

\bibitem{GRW}
G.C. Ghirardi,  A. Rimini, and T. Weber,
 Unified dynamics for microscopic and macroscopic systems,
 Phys. Rev. D 34, 470 (1986).


\bibitem{Bell2}
J. Bell,
Are there quantum jumps?
in {\it Schr\"odinger, Centenary  of a Polymath}, C.W. Kilmister, ed., Cambridge University Press, Cambridge (1987).

\bibitem{Pear2}
P. Pearle,
Combining stochastic dynamical statevector reduction with spontaneous localization,
Phys. Rev. A 39, 2277 (1989).

\bibitem{GirarPearle}
G.C. Ghirardi, P. Pearle,  and A. Rimini,
Markov-processes in Hilbert-space and continuous spontaneous localization of systems of identical particles,
 Phys. Rev. A 42,  78 (1990).


\bibitem{Dio}
L. Diosi,
A universal master equation for the gravitational violation of quantum mechanics,
 Phys. Lett. A, 120, 377 (1987).


\bibitem{Penr}
R.Penrose,
On gravity's role in quantum state reduction,
 Gen. Rel. Grav. 28, 581 (1996).

\bibitem{BDH}
A. Bassi, D. D\"urr, G. Hinrichs,
Uniqueness of the equation for quantum state vector collapse,
 Phys. Rev. Lett. 111, 210401 (2013).

\bibitem{PearleY80}
P. Pearle,
  Collapse miscellany,  in {\it Quantum Theory: A Two-Time Success Story},   D.C. Struppa and J. M. Tollakson, eds. (Springer, New York), p. 131 (2013).

\bibitem{AlV}
D.Z. Albert and L. Vaidman,
 On a proposed postulate of state-reduction,
Phys. Lett. A 139, 1 (1989).


\bibitem{GRWanswer}
G.C. Ghirardi, R. Grassi, F. Benatti,
Describing the macroscopic world: closing the circle within the dynamical reduction program,
Found.  Phys. 25, 5 (1995).

\bibitem{tail1}
P. Lewis,
GRW and the tails problem,
 Topoi 14, 23  (1995).

\bibitem{tail2}
D. Z. Albert, B. Loewer,
Tails of Schr\"odinger's cat,
in {\it Perspectives on Quantum Reality},
The University of Western Ontario Series in Philosophy of Science, Vol. 57, p. 81 (1996).



\bibitem{tail3}
A. Cordero,
Are GRW tails as bad as they say?
Phil. Sci. 66, S59 (1999)


\bibitem{Pear3}
P. Pearle,
Experimental tests of dynamical state-vector reduction,
Phys. Rev. D 29, 235 (1984).


\bibitem{Adl}
S.L. Adler,
Lower and upper bounds on CSL parameters from latent image formation and IGM heating,
J. Phys. A 40, 2935 (2007).


\bibitem{Col95}
B. Collett,  P. Pearle, F. Avignone, and S. Nussinov,
Constraint on collapse models by limit on spontaneous X-ray emission in Ge,
Found. Phys. 25, 1399  (1995).

\bibitem{Bowm}
W.C. Marshall, R. Simon, R. Penrose, and D. Bouwmeester,
Towards quantum superpositions of a mirror,
Phys. Rev. Lett. 91, 130401 (2003).

\bibitem{Arndt}
S. Nimmrichter, K. Hornberger, P. Haslinger, and M. Arndt,
Testing spontaneous localization theories with matter-wave interferometry,
Phys. Rev. A 83, 043621 (2011).



\bibitem{Bassi}
A. Bassi, K. Lochan, S. Satin, T.P. Singh, and H. Ulbricht,
Models of wave-function collapse, underlying theories, and experimental tests,
Rev. Mod. Phys. 85, 471 (2013).

\bibitem{GRWm}
D. Bedingham, D. D\"urr, G.C. Ghirardi, S. Goldstein, R. Tumulka, and N. Zangh\`i,
Matter density and relativistic models of wave function collapse,
J.  Stat. Phys. 154, 623 (2014).


\bibitem{Tumu}
R. Tumulka,
A relativistic version of the Ghirardi-Rimini-Weber model,
J. Stat. Phys. 125,  825 (2006).


\bibitem{EsfGis}
M. Esfeld and N. Gisin,
The GRW flash theory: a relativistic quantum ontology of matter in space-time?
Phil. Sci. 81,  248 (2014).

\bibitem{AlbertWF}
D. Albert,
 Wave function realism,
in {\it The Wave Function: Essays in the Metaphysics of Quantum Mechanics}, edited by D. Albert and A. Ney, p. 52  Oxford University Press, (2013).

\bibitem{GoldZang}
S. Goldstein and  N. Zangh\`i,
Reality and the role of the wave function in quantum theory,
 in {\it The Wave Function: Essays in the Metaphysics of Quantum Mechanics}, edited by D. Albert and A. Ney, p. 91  Oxford University Press, (2013).



\bibitem{Maudlin}
T. Maudlin,
Can the world be only wavefunction?
 in {\it Many Worlds? Everett, Quantum Theory, \& Reality,} S. Saunders, J. Barrett, A. Kent and D. Wallace, eds.,  Oxford and New York: Oxford University Press, p. 121  (2010).

\bibitem{ABL}
Y. Aharonov, P. G. Bergmann, and J. L. Lebowitz,
 Time symmetry in the quantum process of measurement,
  Phys. Rev. 134, B1410 (1964).


\bibitem{AGrus}
Y. Aharonov, E.Y. Gruss,
Two-time interpretation of quantum mechanics,
arXiv:quant-ph/0507269.



\bibitem{CohAha}
Y. Aharonov, E. Cohen, E. Gruss, and  T. Landsberger,
Measurement and collapse within the Two-State-Vector formalism,
Quantum Stud.: Math. Found. ??? (2014).

\bibitem{AV90}
Y. Aharonov  and L. Vaidman,
Properties of a quantum system during the time interval between two measurements,
Phys. Rev. A 41, 11 (1990).

\bibitem{tisy}
  L. Vaidman,
  Time symmetry and the Many-Worlds interpretation,
{\it Many Worlds? Everett, Quantum Theory, \& Reality,} S. Saunders, J. Barrett, A. Kent and D. Wallace, eds.,  Oxford and New York: Oxford University Press, p. 582  (2010).


\bibitem{Broglie28}
L.~de~Broglie,
 La nouvelle dynamique des quanta, in {\it Electrons et photons: rapports et discussions du cinquieme conseil de physique}, Gauthier-Villars,
Paris,  p. 105 (1928);
  English translation:  The new dynamics of quanta, in G. Bacciagaluppi and A. Valentini, {\it Quantum theory at the crossroads: reconsidering the 1927 Solvay conference,}
 Cambridge U.P., Cambridge, (2009).


\bibitem{deBroglie1956}
L.~de~Broglie,
Une tentative d'interpretation causale et non lin{\'e}aire de
  la m{\'e}canique ondulatoire (la th{\'e}orie de la double solution),
Gauthier-Villars, Paris, (1956);
 English translation:   Non-linear wave mechanics: A Causal
  Interpretation, Elsevier, Amsterdam, (1960).

\bibitem{Bohm52}
D. Bohm,
 A suggested interpretation of the quantum theory in terms of ``hidden''
  variables, I and II,
 Phys. Rev.   85,  166  (1952).

\bibitem{bell2004speakable}
J.S.~Bell,
 de Broglie-Bohm, delayed-choice double-slit experiemnt, and density matrix,
 Int. J. Quan. Chem.  14,  155 (1980).

\bibitem{Golds}
D. D\"urr, S. Goldstein,   and  N. Zangh\`i,
Quantum Equilibrium and the Origin of Absolute Uncertainty,
J. Stat. Phys. 67, 843 (1992).

\bibitem{PhysRev.89.458}
D.~Bohm,
  Proof that probability density approaches $|\psi|^{2}$ in causal   interpretation of the quantum theory,
  Phys. Rev.  89,  458 (1953).

\bibitem{Valentini2005}
A. Valentini and H. Westman,
  Dynamical origin of quantum probabilities,
  Proc. R. Soc. A   461, 253 (2005).

\bibitem{Valentini2010}
A. Valentini,
Inflationary cosmology as a probe of primordial quantum mechanics,
Phys. Rev. D 82, 063513 (2010).

\bibitem{gili}
G. Naaman-Marom, N. Erez, and L. Vaidman,
Position measurements in the de Broglie-Bohm interpretation of quantum mechanics,
 Ann. Phys. 327, 2522 (2012).


\bibitem{Ghirardi}
G.~C. Ghirardi,
 {\it Sneaking a look at god cards},
Princeton University Press, Princeton  p. 213 (2004).


\bibitem {BohmSpin}
C. Dewdney, P.R. Holland, A. Kyprianidis, and J.P. Vigier,
Spin and non-locality in quantum mechanics,
Nature,  336, 536 (1988).


\bibitem{Goldstein}
D.~D\"urr, S.~Goldstein, and N. Zangh\`i,
  Quantum equilibrium and the role of operators as observables in  quantum theory,
  J. Stat. Phys.  116,  959 (2004).

\bibitem{BellHid}
J.S.~Bell,
On the problem of hidden variables in quantum mechanics,
Rev. Mod. Phys. 38, 447 (1966).


\bibitem{Gold2}
D.~D\"urr, S.~Goldstein, T. Norsen, W. Struyve, and N. Zangh\`i,
 Can Bohmian mechanics be made relativistic?
 Proc. Roy. Soc. A 470,  20130699 (2013).


\bibitem{Englert}
B.G. Englert, M.~O.~Scully, G.~S\"{u}ssmann, and H.~Walther,
 Surrealistic Bohm trajectories,
 Z. Naturforsch. A  47,  1175 (1992).

\bibitem{Tipler2006}
F.J. Tipler,
 What about quantum theory? Bayes and the Born interpretation,
 arXiv:quant-ph/0611245 (2006).


\bibitem{Val2010}
A. Valentini,
 De Broglie-Bohm pilot-wave theory: Many Worlds in denial?
 {\it Many Worlds? Everett, Quantum Theory, \& Reality,} S. Saunders, J. Barrett, A. Kent and D. Wallace, eds.,  Oxford and New York: Oxford University Press, p. 121  (2010).

\bibitem{Bo}
K.J. Bostr\"om,
 Combining Bohm and Everett: Axiomatics for a standalone quantum mechanics,
  arXiv:1208.5632 (2012).

\bibitem{Hall}
M.J.W. Hall, D.-A. Deckert, H.M. Wiseman,
Quantum phenomena modelled by interactions between many classical worlds,
arXiv:1402.6144  (2014).

\bibitem{Se}
C.T. Sebens,
Quantum mechanics as classical physics,
arXiv:1403.0014 (2014).

\bibitem{AlLo88}
D.  Albert and B. Loewer,
Interpreting the many worlds interpretation,
Synthese 77, 195 (1988).


\bibitem{Eve}
 H. Everett,
 Relative state formulation of quantum mechanics,
  Rev.  Mod. Phys. 29, 454 (1957).

 \bibitem{Tip86}
F. J. Tipler,
 The Many-Worlds interpretation of quantum mechanics in quantum cosmology,
  in  {\it Quantum Concepts of Space and Time}, R. Penrose and C.J. Isham eds.,  Oxford: The Clarendon Press   p. 204 (1986).


\bibitem{AgTeg}
A.  Aguirre, and M. Tegmark,
Born in an infinite Universe: A cosmological interpretation of quantum mechanics,
 Phys. Rev. D 84, 105002  (2011).

 \bibitem{DeJo}
D.  Deutsch,  and  R. Jozsa,
Rapid solutions of problems by quantum computation,
 Proc. Roy. Soc.  London A 439, 553  (1992).


\bibitem{V-Pit}
  L. Vaidman,
  Probability in the Many-Worlds interpretation of quantum mechanics,
in {\it Probability in Physics}, The Frontiers Collection, Y. Ben-Menahem and M. Hemmo, eds.,  p. 299  Springer-Verlag, Berlin Heidelberg (2012).



\bibitem{Zur}
W.H. Zurek,
Decoherence, einselection, and the quantum origins of the classical,
Rev. Mod. Phys. 75, 715 (2003).


\bibitem{Schlos}
M. Schlosshauer,
{\it Decoherence and the quantum-to-classical transition},
Heidelberg and Berlin: Springer (2007).

\bibitem{Zeh}
H.D. Zeh,
Roots and fruits of decoherence,
in {\it Quantum Decoherence},
Prog. Math. Phys.   48, 151 (2007).

 \bibitem{Wallace}
D. Wallace,
 {  \it The Emergent Multiverse:  Quantum Theory according to the Everett Interpretation},
 Oxford: Oxford University Press  (2012).

\bibitem{Wa03}
D. Wallace,
 Everettian rationality: Defending Deutsch's approach to probability in the Everett interpretation,
Stud. Hist. Phil. Mod. Phys. 34, 415 (2003).

\bibitem{schizo}
L. Vaidman,
 On schizophrenic experiences of the neutron or why we should believe in the Many-Worlds interpretation of quantum theory,
  Int. Stud. Phil. Sci. 12, 245 (1998).



\bibitem{TUWS}
{\it Tel Aviv Quantum World Splitter},
http://qol.tau.ac.il/

\bibitem{Al10}
D. Albert,
Probability in the Everett picture,
 in {\it Many Worlds? Everett, Quantum Theory, \& Reality,} S. Saunders, J. Barrett, A. Kent and D. Wallace, eds.,  Oxford and New York: Oxford University Press, p. 355  (2010).

\bibitem{SaWa2008a}
S. Saunders,  and D. Wallace,
Branching and uncertainty,
 Br. J. Philos.  Sci.  59, 293 (2008).

\bibitem{Ta2011}
P. Tappenden,
Evidence and uncertainty in Everett's multiverse,
  Br. J. Philos.  Sci.  62, 99 (2011).

\bibitem{Deut}
D. Deutsch,
Quantum theory of probability and decisions,
Proc. Roy. Soc. Lon. A 455, 3129 (1999).


\bibitem{Wala}
D. Wallace,
Quantum probability from subjective likelihood: Improving on Deutsch's proof of the probability rule,
 Stud. Hist. Phil. Mod. Phys. 38, 311 (2007).

\bibitem{Elga}
A. Elga,
 Self-locating belief and the Sleeping Beauty problem,
Analysis 60, 143  (2000).

\bibitem{Lewis}
D. Lewis,
Sleeping Beauty: reply to Elga,
Analysis   61, 171 (2001).

\bibitem{SBAna}
B. Groisman, N. Hallakoun, and L. Vaidman,
 The measure of existence of a quantum world and the Sleeping Beauty Problem,
 Analysis 73, 695  (2013).

\bibitem{PeLew}
P.J. Lewis,  Quantum Sleeping Beauty,
 Analysis 67, 59 (2007).

\bibitem{AllGT}
V. Allori, S. Goldstein, R. Tumulka, and N. Zangh\`i,
 Predictions and primitive ontology in quantum foundations: A study of examples,
Br. J. Philos. Sci.  65, 323 (2014).


\bibitem{AB}
Y. Aharonov and D. Bohm,
Significance of electromagnetic potentials in the quantum theory,
Phys. Rev. 115, 485 (1959).


\bibitem{VAB}
L. Vaidman,
Role of potentials in the Aharonov-Bohm effect,
Phys. Rev. A 86, 040101(R) (2012).

\bibitem{Tirzamod}
T. Kaufherr, Y. Aharonov, S. Nussinov, S. Popescu, and J. Tollaksen,
Dynamical features of interference phenomena in the presence of entanglement,
Phys. Rev. A 83, 052127 (2011).




\bibitem{EV}
A. C.  Elitzur and L. Vaidman,
Quantum mechanical interaction-free measurements,
Found. Phys.  23, 987 (1993).

\bibitem{IFMzeno}
P. G. Kwiat, A. G. White, J. R. Mitchell, O. Nairz, G. Weihs, H. Weinfurter, and A. Zeilinger,
High-efficiency quantum interrogation measurements via the Quantum Zeno Effect,
Phys. Rev. Lett. 83, 4725 (1999).

\bibitem{Jozsa}
R. Jozsa,
Quantum effects in algorithms,
 in {\it Quantum Computing and Quantum Communications}, Lecture Notes in Computer Science,
C. P. Williams, ed. , Vol. 1509, p. 103 Springer, London, (1999).

\bibitem{Noh}
T.-G. Noh,
Counterfactual quantum cryptography,
Phys. Rev. Lett. 103, 230501 (2009).


\bibitem{exp}
Y. Liu, L. Ju, X.-L. Liang, S.-B. Tang, G.-L. S. Tu, L. Zhou, C.-Z. Peng, K. Chen, T.-Y. Chen, Z.-B. Chen, and J.-W. Pan,
Experimental demonstration of counterfactual quantum communication,
Phys. Rev. Lett. 109, 030501 (2012).


\bibitem{Zub}
H. Salih, Z.H. Li, M. Al-Amri, and M.S. Zubairy,
Phys.  Rev. Lett. {\bf 110}, 170502  (2013).
Protocol for direct counterfactual quantum communication,
H. Salih, Z.-H. Li, M. Al-Amri, and M.S. Zubairy,
Phys. Rev. Lett. 110, 170502 (2013)

\bibitem{Cao}
Y. Cao, Y.-H. Li, Z. Cao, J. Yin,. Y.-A. Chen, X. Ma,. C.-Z. Peng, and J.-W. Pan.,
Direct counterfactual communication with single photons,
arXiv:1403.5082.

\bibitem{Guo}
Q. Guo, L.-Y. Cheng, L. Chen, H.F. Wang, and S. Zhang,
Counterfactual entanglement distribution without transmitting any particles,
Opt. Expr. 22,   8970 (2014).

\bibitem{Salih}
H. Salih,
Protocol for counterfactually transporting an unknown qubit,
 arXiv:1404.2200.

\bibitem{VcomSalih}
 L. Vaidman,
Comment on ``Protocol for direct counterfactual quantum communication'',   Phys. Rev. Lett. 112, 208901 (2014).

\bibitem{Danan}
A. Danan, D. Farfurnik, S. Bar-Ad, and L. Vaidman,
Asking photons where they have been,
Phys. Rev. Lett. 111, 240402 (2013).

\bibitem{RVCom}
H. Salih, Z.H. Li, M. Al-Amri, and M.S. Zubairy,
Reply to Comment on ``Protocol for direct counterfactual quantum communication'',
 Phys. Rev. Lett. 112, 208902 (2014).

\bibitem{ComPast}
Z.H. Li, M. Al-Amri, and M.S. Zubairy,
Comment on ``Past of a quantum particle''
Phys. Rev. A 88, 046102 (2013).

\bibitem{RComPast}
L. Vaidman,
 Reply to the ``Comment on {`Past of a quantum particle'} '',
Phys. Rev. A 88 , 046103 (2013).

\bibitem{Vpast}
L. Vaidman,
Past of a quantum particle,
Phys.  Rev. A  87, 052104 (2013).


\end{thebibliography}
\end{document}